%% file: main.tex
\documentclass[12pt]{article}
\usepackage{xr-hyper}
\usepackage[hidelinks]{hyperref}
\usepackage{arxiv}
\usepackage[english]{babel}
\usepackage[utf8]{inputenc}
\usepackage{array}
\usepackage{booktabs}
\usepackage{caption}
\usepackage[T1]{fontenc}    
\usepackage{microtype}      
\newcommand{\tocline}[2]{%
  \hyperref[#1]{#2}\dotfill\pageref*{#1}%
}

\usepackage{url}
\usepackage{makecell}

\usepackage{nicefrac}       
\usepackage{doi}

\usepackage{amsmath,amsfonts,amssymb,amscd,amsthm,xspace}
\usepackage{graphicx}
\usepackage[numbers,sort&compress]{natbib}
\usepackage[colorinlistoftodos,
            tickmarkheight=.5em,
            textwidth=\marginparwidth, 
            textsize=small]{todonotes}

\usepackage[export]{adjustbox}
\usepackage{enumitem}
\usepackage{multirow}
\usepackage{setspace} 
\usepackage{colortbl}
\definecolor{lightgray}{gray}{0.9}

\usepackage{algorithm}
\usepackage[indLines, 
            noEnd=false,
            beginComment=$\#$~,
            beginLComment=$\#$~,
            endLComment=,
            ]{algpseudocodex}

\usepackage{listings}
\definecolor{codegray}{rgb}{0.5,0.5,0.5}
\definecolor{backcolour}{rgb}{0.95,0.95,0.92}
\newcolumntype{P}[1]{>{\centering\arraybackslash}p{#1}}
\lstdefinestyle{mystyle}{
    backgroundcolor=\color{backcolour},   
    numberstyle=\tiny\color{codegray},
    basicstyle=\ttfamily\footnotesize,
    breakatwhitespace=false,
    breaklines=true,
    captionpos=b,
    keepspaces=false,
    numbers=left,
    numbersep=5pt,
    showspaces=false,
    showstringspaces=false,
    showtabs=false,
    tabsize=2
}
\title{Bayesian Optimization of Catalysis with In-Context Learning}
\date{\today}

\author{
    \href{https://orcid.org/0000-0001-5336-2847}{\includegraphics[scale=0.06]{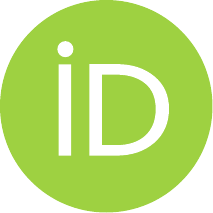}\hspace{1mm}Mayk Caldas Ramos\textsuperscript{ *}}\\
	FutureHouse Inc., San Francisco, CA\\
	Department of Chemical Engineering \\
	University of Rochester\\
	\texttt{mayk@futurehouse.org} \\
\And
    \href{https://orcid.org/}{\includegraphics[scale=0.06]{orcid.pdf}\hspace{1mm}Shane S. Michtavy\textsuperscript{ *}}\\Department of Chemical Engineering \\
	University of Rochester\\
	\texttt{smichtav@che.rochester.edu} \\
\And
	\href{https://orcid.org/0000-0003-3066-0029}{\includegraphics[scale=0.06]{orcid.pdf}\hspace{1mm}Marc D. Porosoff\textsuperscript{ \dag}}\\
	Department of Chemical Engineering \\
	University of Rochester\\
	\texttt{marc.porosoff@rochester.edu} \\
\And
	\href{https://orcid.org/0000-0002-6647-3965}{\includegraphics[scale=0.06]{orcid.pdf}\hspace{1mm}Andrew D. White\textsuperscript{ \dag}}\\
    FutureHouse Inc., San Francisco, CA\\
	Department of Chemical Engineering \\
	University of Rochester\\
	\texttt{andrew@futurehouse.org} \\
}

\renewcommand{\headeright}{Article}
\renewcommand{\undertitle}{Article}
\renewcommand{\shorttitle}{Bayesian Optimization of Catalysis with In-Context Learning}

\hypersetup{
pdftitle={Bayesian Optimization of Catalysis With In-Context Learning},
pdfsubject={},
pdfauthor={Ramos, Mayk and Michtavy, Shane and Porosoff, Marc and White, Andrew D},
pdfkeywords={Bayesian Optimization, large language model, LLM, in-context learning, catalysis, materials design, AI},
}

\begin{document}
\maketitle

\begingroup
\renewcommand\thefootnote{}\footnotetext{\textsuperscript{*}These authors contributed equally to this work.}
\footnotetext{\textsuperscript{\dag}Corresponding authors.}
\endgroup

\begin{abstract}
Large language models (LLMs) can perform accurate classification with zero or few examples through in-context learning. We extend this capability to regression with uncertainty estimation using frozen LLMs (e.g., GPT-3.5, Gemini), enabling Bayesian optimization (BO) in natural language without explicit model training or feature engineering. We apply this to materials discovery by representing experimental catalyst synthesis and testing procedures as natural language prompts.

A key challenge in materials discovery is the need to characterize suboptimal candidates, which slows progress. While BO is effective for navigating large design spaces, standard surrogate models like Gaussian processes assume smoothness and continuity, an assumption that fails in highly non-linear domains such as heterogeneous catalysis. Our task-agnostic BO workflow overcomes this by operating directly in language space, producing interpretable and actionable predictions without requiring structural or electronic descriptors.

On benchmarks like aqueous solubility and oxidative coupling of methane (OCM), BO-ICL matches or outperforms Gaussian processes. In live experiments on the reverse water-gas shift (RWGS) reaction, BO-ICL identifies near-optimal multi-metallic catalysts within six iterations from a pool of 3,700 candidates. Our method redefines materials representation and accelerates discovery, with broad applications across catalysis, materials science, and AI. Code: \url{https://github.com/ur-whitelab/BO-ICL}.
\end{abstract}

\input{paper}

\subsubsection*{Acknowledgments}
This research is supported by the National Science Foundation under Grant No. CBET-2345734, the National Institute of General Medical Sciences of the National Institutes of Health (NIH) under award number R35GM137966, and the U.S. Department of Energy, Grant No. DE-SC0023354.
The authors also thank the computational resources and structure provided by the Center for Integrated Research Computing (CIRC) at the University of Rochester, as well as Rashad Ahmadov for help with catalyst synthesis preparations.

\bibliographystyle{unsrtnat}
\bibliography{refs}
\clearpage          
\thispagestyle{empty}

\setcounter{section}{0}
\setcounter{figure}{0}
\setcounter{table}{0}
\setcounter{page}{1}

\renewcommand\thesection   {S\arabic{section}}
\renewcommand\thesubsection{S\arabic{section}.\arabic{subsection}}
\renewcommand\thefigure    {S\arabic{figure}}
\renewcommand\thetable     {S\arabic{table}}

\makeatletter
\newcommand{\SIsection}[1]{%
  \refstepcounter{section}
  \section*{\thesection\quad #1}
  \addcontentsline{toc}{section}{S\thesection\qquad #1}}
\makeatother

\begin{center}
\textbf{\LARGE Supporting Information for: \\[0.2em]
Bayesian Optimization of Catalysis with In-Context Learning}\\[.5cm]

\begin{tabular}{P{6cm}P{6cm}}
\textbf{Mayk Caldas Ramos}$^{1,2,*}$ & \textbf{Shane S. Michtavy}$^{1,*}$\\
\texttt{mayk@futurehouse.org}        & \texttt{smichtav@che.rochester.edu}\\[.6em]

\textbf{Marc D. Porosoff}$^{1,\dag}$ & \textbf{Andrew D. White}$^{1,2,\dag}$\\
\texttt{marc.porosoff@rochester.edu} & \texttt{andrew@futurehouse.org}
\end{tabular}

\vspace{0.5cm}
$^{1}$ Department of Chemical Engineering, University of Rochester\\
$^{2}$ FutureHouse Inc., San Francisco, CA\\[0.5em]
$^{*}$ These authors contributed equally to this work.\\
$^{\dag}$ Corresponding authors
\end{center}

\vfill

\section*{\centering Contents of Supporting Information}
\addcontentsline{toc}{section}{Contents of Supporting Information}

\begin{center}
\begin{minipage}{0.85\linewidth}
\begin{enumerate}[leftmargin=2em,label=\textbf{S\arabic*.}]
  \item \tocline{sec:icl_algs}{BO-ICL Algorithms}
  \item \tocline{sec:cost-analysis}{Cost Analysis}
  \item \tocline{sec:sysmes}{Prompts and System Messages}
  \item \tocline{sec:datasets}{Datasets}
    \begin{enumerate}[label=\textbf{S\arabic{enumi}.\arabic*.}]
      \item \tocline{ssec:solubility}{Solubility}
      \item \tocline{ssec:ocm}{Oxidative Coupling of Methane}
      \item \tocline{ssec:alloy}{Alloy Interface}
      \item \tocline{ssec:rwgs}{In-house RWGS}
    \end{enumerate}
  \item \tocline{sec:baselines}{Baselines}
    \begin{enumerate}[label=\textbf{S\arabic{enumi}.\arabic*.}]
      \item \tocline{ssec:an_rand}{Analytical Random}
      \item \tocline{ssec:rand_label}{OCM with No True Correlation}
      \item \tocline{ssec:knn}{k-Nearest Neighbor}
      \item \tocline{ssec:krr}{Kernel Ridge Regression}
      \item \tocline{ssec:gpr}{Gaussian Process}
    \end{enumerate}
  \item \tocline{sec:reg_results}{Additional Results}
    \begin{enumerate}[label=\textbf{S\arabic{enumi}.\arabic*.}]
      \item \tocline{sec:solubility}{Solubility}
      \item \tocline{sec:regocm}{Regression – OCM}
      \item \tocline{sec:mmr}{MMR Dependence}
    \end{enumerate}
\end{enumerate}
\end{minipage}
\end{center}

\vfill
\clearpage

\vfill                

\clearpage            

\input{supMat}

\end{document}

%% file: paper.tex
\begin{figure}[hbt!]
    \centering
    \includegraphics[width=1.0\linewidth]{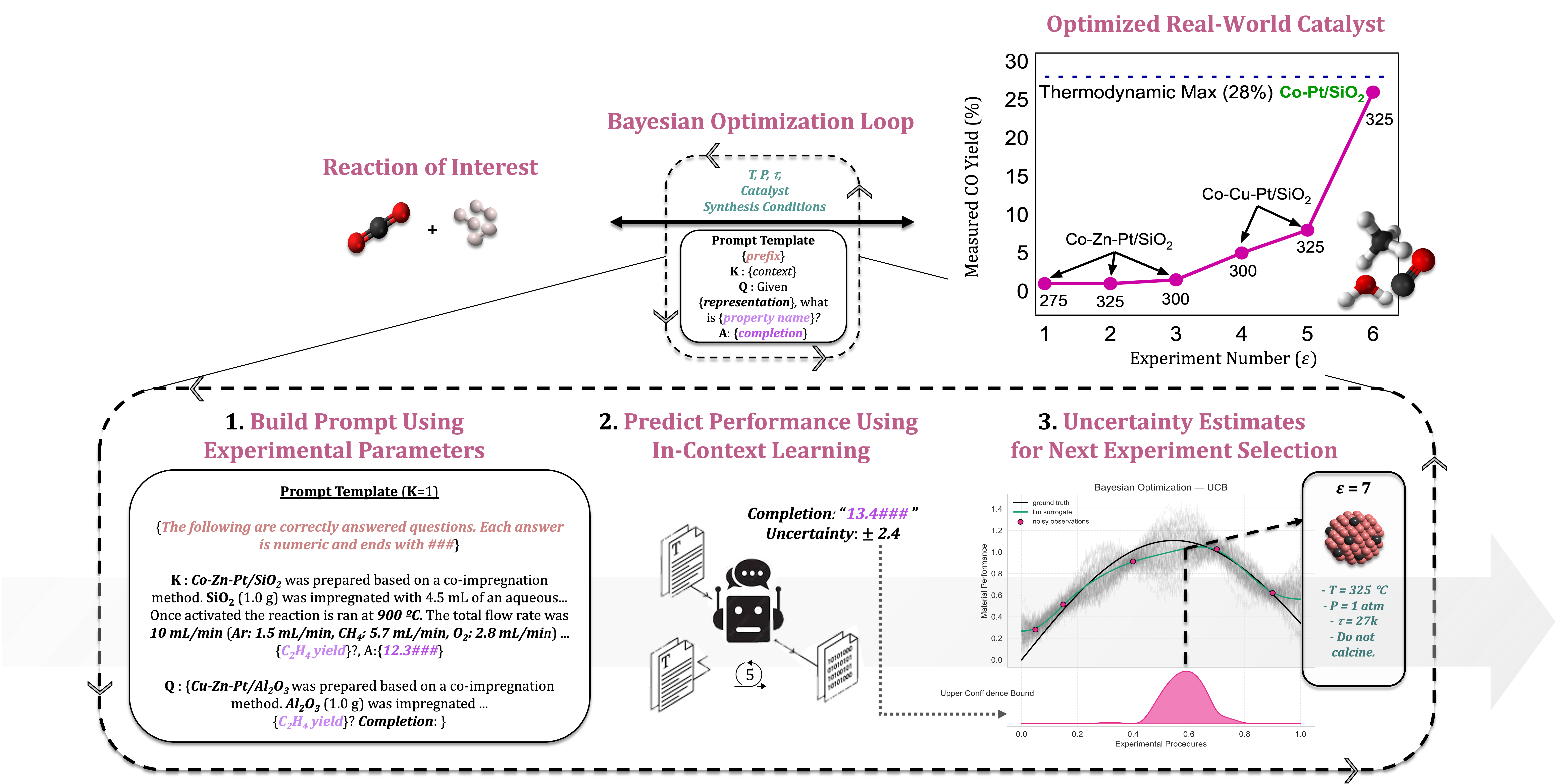}
    \addtocounter{figure}{0}
    \caption{A high-level overview of a closed-loop Bayesian optimization (BO) method that uses natural language to represent a material design space for efficient sample space exploration. The workflow involves conversion of tabular data into an experimental procedure, which incorporates both synthesis and reaction parameters. By formatting material parameters for compatibility with state-of-the-art large language models, this approach leverages well-established BO techniques to efficiently identify actionable experimental conditions that maximize a desired objective function. In this figure, we highlight a success case for optimizing catalysts for selective $CO_2$ conversion to $CO$ via BO-ICL}
    \label{fig:toc}
\end{figure}

\section{Introduction}

Transformer large language models (LLMs) have impacted a range of domains because of their task-agnostic training process\cite{Vaswani2017-az}, where the same pre-training process can be applied to obtain state-of-the art models across many scientific domains \cite{Rajpurkar2018-du, Sun2019-vh, Chen2021-ec, He2023-qr, white2023assessment, Kung2023-wm, Hassan2024-sq, Lu2024-xl}.
LLMs are applicable beyond natural language applications and include fields such as medicine\cite{Coppersmith2014-hh, Schwaller2021-qr, Schwaller2022-lu,Khader2023-ec,Gwon2024-mn}, materials property prediction\cite{Blanchard2022-vp, Xu2022-ti, Ross2022-na, Jablonka2023-ya, Yu2024-gr, Landram2024-ae}, and molecular design\cite{Honda2019-jb, Ozturk2020-cd, Liu2021-ds, Wellawatte2025-fz, Suvarna2023-ue, Ramos2025-pg}. Their ability to gain accuracy via in-context learning (ICL), whereby one to five examples can improve accuracy, is also remarkably unique among modeling approaches.\cite{Brown2020-bf} In this work, we explore if the ICL property of LLMs can be combined with optimization to design specific new materials.

Bayesian optimization (BO) is a common technique for constrained optimization applications\cite{Frazier2018-hq}.
BO addresses the problem:
\begin{equation}
\arg\max_{x \in \Omega} f(x)
\end{equation}
which translates to \emph{finding the input $x$ within the design space $\Omega$ that maximizes the objective function $f$}.
BO often uses predictions and uncertainty estimates from probabilistic models to efficiently balance exploration and exploitation in the search for optimal parameters\cite{Lookman2017-pm, Liang2021, hernandez2017predictive, eyke2020iterative}.
More specifically, BO performs gradient-free optimization of a black-box function $f(x)$ by employing a surrogate model $\mathcal{S}(x)$ to approximate $f(x)$ and an acquisition function $\alpha(x)$ to determine the next evaluation point. As new information about $f(x)$ becomes available, the surrogate model is updated according to the acquisition policy\cite{Lei2021-jp}.
A detailed description of BO is available in Section~\ref{sec:methods}.

A common choice for a surrogate function is a Gaussian process (GP) model; GPs do not impose restrictive parametric assumptions and are inherently probabilistic\cite{Lei2021-jp, Wang2023-oa}. We propose LLMs in this work. LLMs predict probability distributions over their vocabulary, enabling the direct extraction of uncertainty estimations. The combination of available confidence scores and the efficacy of LLMs with ICL supports the idea that these models may be useful for the rapid updates required in BO.

Using an LLM as a surrogate model enables the use of natural language as feature vectors. This is particularly valuable for domain applications that are challenging to model, such as experimental protocols that represent a catalyst\cite{Hegde2017-go,del-Rio2023-rz,Uhrin2025-xw,Ito2024-nv}. Natural language provides a straightforward way to integrate both relevant qualitative and quantitative information into representations, which can then be optimized. Building on this capability, \citet{Jablonka2023-ya} demonstrated that decoder-only models like the generative pretrained transformer (GPT) can predict material and chemical properties using Language-Interfaced Fine-Tuning (LIFT)\cite{Dinh2022-oj,Brown2020-bf}.
LIFT converts tabular data into sentences and then fine-tunes an LLM using the resulting natural language representation (similar to the illustration in Figure \ref{fig:tab_text} of the SI). 

The application of LIFT using GPT models has succeeded in tasks such as classification, regression, and inverse design, without requiring modifications to model architectures or training procedures\cite{Dinh2022-oj, Jablonka2024-wq}. However, using GPT models as surrogates for BO introduces additional challenges, such as the requirement of substantially more training compute. Surrogate models are updated upon each observation in BO, which will be a significant additional burden on LLM training in the LIFT paradigm\cite{Radford2019LanguageMA}.

Fortunately, there are alternative strategies to re-training the LLM upon a BO update, such as ICL.\cite{Monea2024-av}. ICL enhances performance by allowing the model to observe query-relevant examples at inference time \cite{Brown2020-bf}, eliminating the need for additional weight updates to generalize beyond its original training data \cite{Su2022-bq, Xu2023-ik}. 
Recent research highlights success using similar ICL prompting techniques, such as chain-of-thought \cite{wei2022chain, kojima2022large, jung2022maieutic} and the use symbolic tools (e.g., programming languages) to improve accuracy\cite{zhou2022teaching, lyu2023faithful}. 
Thus, ICL enables models to improve prediction accuracy even when new data is available at a limited rate, a useful attribute for a BO workflow.

The integration of pre-trained LLMs with BO has become an active area of research after our early demonstrations of their potential \cite{Liu2024-mq,Gwon2024-mn}. Notably, \citet{Kristiadi2024-mf} shows that using domain-specific LLMs, trained via parameter-efficient fine-tuning (PEFT), achieve success in simpler BO settings. 
With inspiration from these prior ideas, we present a novel approach that successfully leverages LLMs as surrogate models in a BO policy via ICL. Figure~\ref{fig:toc} shows a high-level illustration of our method of integrating BO with ICL, and further details are available in Figure ~\ref{fig:flowchart}. Our process introduces an AskTell algorithm that utilizes ICL as the primary mechanism for updating the surrogate LLM's knowledge during the BO process. 

AskTell means we first query the model for a point with an Ask and then we respond to the model with a Tell step, reporting the outcome of the experiment. By dynamically constructing prompts with relevant context at inference time, we eliminate the need for resource intensive weight updates, as is common when updating a model as new data becomes available. This yields a task-agnostic, ready-to-use approach that operates directly in natural language space.

To validate our workflow, we focus on materials design for greenhouse gas (GHG) upcycling, an application area of global significance. Accelerating materials discovery in this domain can reduce reliance on crude oil for high-demand precursors such as carbon monoxide and olefins \cite{Juneau2020-bi}. By targeting heterogeneous catalytic reactions involving GHGs such as $CO_2$, we may help close the GHG emission life cycle responsible for atmospheric accumulation, thereby mitigating global temperature rise \cite{Wang2021-dn, Duyar2015-gs}. Enhancing materials design and discovery has the potential to impact each step in such a circular carbon economy by helping to offset the inherent entropic penalties associated with the capture and conversion of relevant GHGs \cite{Duyar2015-gs,Ahmadov2024-jb}. 

Given the vast design space of heterogeneous catalysts and the additional complexity of reaction condition optimization, catalysis offers a compelling use case for frozen LLMs as surrogate models within a BO framework\cite{Ahmadov2024-jb}. Language-based representations of materials allow experimentalists to optimize catalytic performance by formatting inputs—such as synthesis procedures and reaction conditions—in a simple, structured way, with property values as outputs (see Fig.~\ref{fig:toc}). Leveraging pre-trained LLMs for prompt-level transfer learning is expected to improve optimization efficiency, reduce experimental overhead, and accelerate catalyst discovery.

In this work, we investigate whether ICL with state-of-the-art GPT models  serves as an effective surrogate model within a BO framework. Our central hypothesis is that language-based representations contain sufficient structure and physical information to enable efficient experimental design, even without domain-specific feature engineering. We begin by evaluating scalability through two regression tasks: predicting molecular solubility from IUPAC names, and catalytic performance in the oxidative coupling of methane (OCM) reaction using natural language descriptions of synthesis and reaction conditions (Section~\ref{sec:regression}). We then assess BO-ICL's sample efficiency on the OCM dataset from \citet{Nguyen2020-kf} and an alloy interface property dataset from \citet{Gerber2023-xl} (Sections~\ref{sec:ocm_bo} and \ref{sec:aii_bo}, respectively), showing rapid convergence to the 1\% top-performing candidates after labeling only thirty experiments. 
Finally, we apply BO-ICL to guide real-world on-the-fly experimental synthesis and testing for the reverse water gas-shift (RWGS) reaction using multi-metallic catalysts, achieving near-thermodynamic equilibrium performance after only six iterations (Section~\ref{sec:rwgs_bo}). 
Together, these results support our goal of enabling general-purpose, language-native optimization workflows for materials design.

\section{Results and discussion}

We use four datasets to evaluate the performance of our method: estimated solubility (ESOL)\cite{Delaney2004-de}, oxidative coupling of methane (OCM)\cite{Nguyen2020-kf},
modeled alloy interface interaction (AII)\cite{Gerber2023-xl}, and an in-house dataset generated for $CO_2$ hydrogenation under RWGS conditions. Detailed descriptions of these datasets are available in Section~\ref{sec:datasets}.

Initially, we employ ESOL and OCM datasets in a regression task to investigate how the performance of our ICL approach depends on key hyperparameters: the number of examples used in the prompt ($k$), the uncertainty scaling factor for calibration, and the temperature ($T$) (see Section~\ref{sec:hyperparam} and Figure~\ref{fig:flowchart}, for use locations). We extend these regression experiments (Section~\ref{sec:regression}) to confirm that the model learns directly from the natural language representations. To benchmark the LLM's performance against other commonly used machine learning models, we test three baseline methods: k-nearest neighbor\cite{altman1992introduction} (knn), kernel ridge regression\cite{Saunders1998-dp, Vu2015-nt} (krr), and Gaussian Process Regression\cite{Rasmussen2005-xi} (GPR). Implementation details for the baselines are provided in Section~\ref{sec:baselines}.

Next, in Section~\ref{sec:bo}, we perform optimization using LLMs as surrogate models combined with ICL to iteratively update model knowledge using the OCM and AII datasets (RAG workflow illustration is in Figure~\ref{fig:flowchart}). We observe that BO-ICL reaches the $99^{\mathrm{th}}$ percentile of active catalysts while requiring, on average, less than thirty new samples.

Finally, we construct an unlabeled pool of potential experiments for in-house synthesis and testing, comprising experimental procedures for the RWGS reaction. We use BO-ICL to iteratively guide the selection of subsequent experiments, with $CO$ yield as the objective function in the RWGS catalyst design space. We demonstrate that BO-ICL effectively selects experimental procedures achieving $CO$ yields closely approaching the thermodynamic limit (see Supporting Information (SI) Section~\ref{sec:datasets}), after only six iterative cycles. All results use an embedded natural language representation of the sampled experimental procedures as the input feature representation.

\subsection{Regression}\label{sec:regression}

We begin our analysis by identifying key hyperparameter values and examining how the number of known examples stored in the model's memory (available context) influences prediction performance using regression analysis (Section~\ref{sec:hyperparam}). Motivated by insights from this exploratory analysis, we conduct subsequent experiments using five context examples per prompt, a temperature setting of $0.7$, and an uncertainty scaling factor of $5$. Figures~\ref{fig:ocm-metrics} and ~\ref{fig:sol-metrics} illustrate the impact of these hyperparameters on prediction performance for the ESOL and OCM datasets using the \texttt{gpt-3.5-turbo-0125} model.

\begin{figure}[h!]
    \centering
    \includegraphics[width=0.7\linewidth]{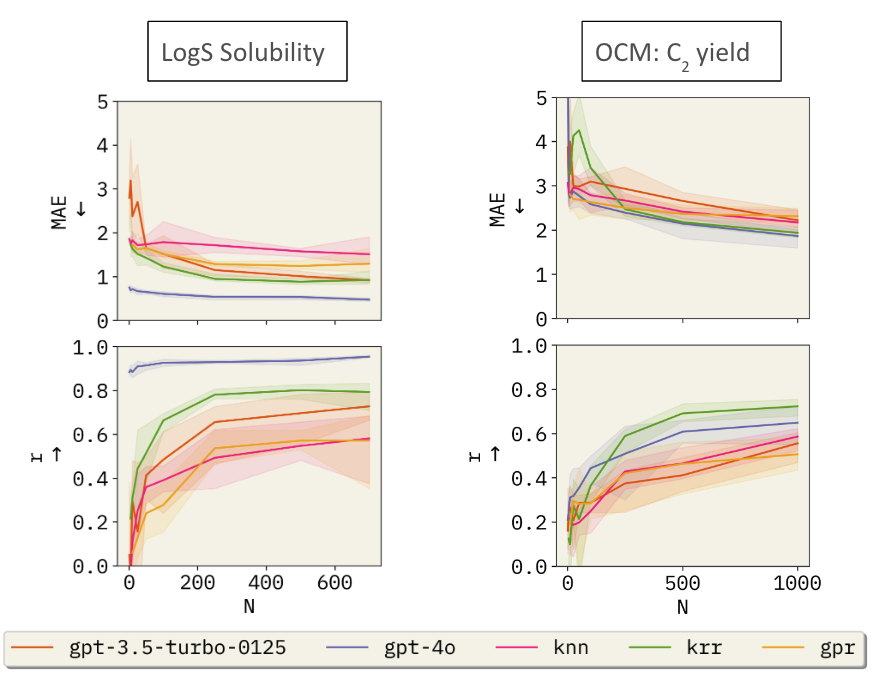}
    \caption{Performance comparison of baseline models versus BO-ICL based on the number of points in the model's memory or used to train, as applicable. The top row shows the Mean Absolute Error (MAE) as a function of the number of training samples (N), while the bottom row shows Pearson correlation (r). The models compared include GPT-3.5-turbo-0125, GPT-4o, Kernel Ridge Regression (KRR), k-Nearest Neighbors (KNN), and a Gaussian Process Regressor. The shaded areas represent the range of the predictions in each replicate.}
    \label{fig:regression-results}
\end{figure}

\begin{figure}[h!]
    \centering
    \includegraphics[width=\linewidth]{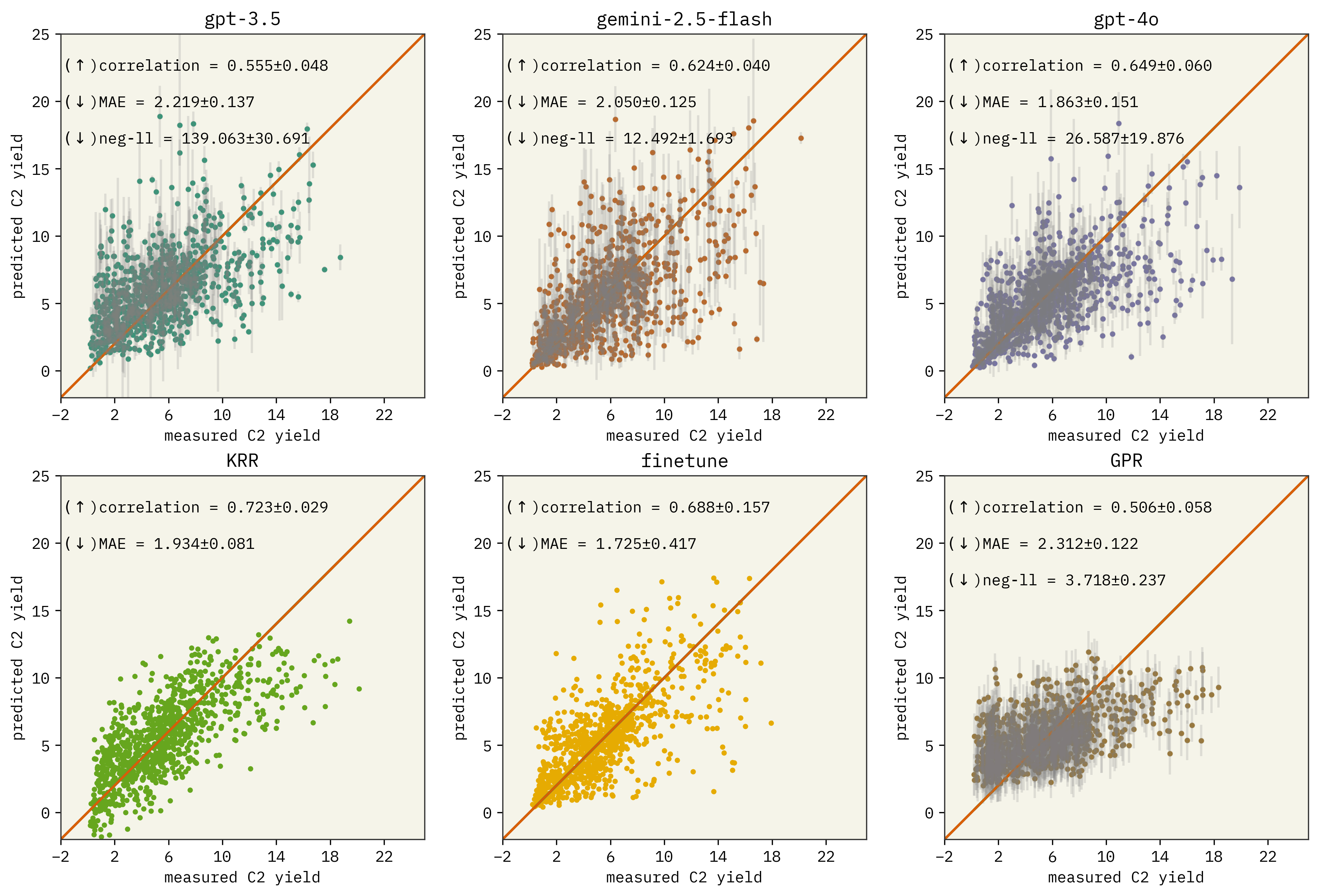}
    \caption{Parity plots for the regression task on the OCM dataset across different models. Each model was evaluated over five independent replicates, with each plot aggregating all predicted vs. true values. Reported metrics reflect the mean and standard deviation across replicates. Large language models (LLMs) exhibit comparable performance, with GPT-4o showing a slight edge. Interestingly, kernel ridge regression (KRR) achieves the highest correlation among all models, though it was not further explored due to its lack of uncertainty estimates.}
    \label{fig:ocm-models}
\end{figure}

To assess the performance of our ICL approach relative to more traditional methods, we benchmark against krr, a fine-tuned variant of \texttt{gpt-3.5-turbo-0125}, and GPR. Figures~\ref{fig:sol-models} and \ref{fig:ocm-models} presents results the solubility and the OCM datasets, respectively. The baselines demonstrate strong performance across datasets in comparison with the ICL approach, consistent with previous findings in the literature \cite{Jablonka2023-ya}. Baseline model performance advantages likely arise from task specific parameter updates, contrasting with the continuous reuse of a single general-purpose LLM in the ICL setup. Specifically, KRR likely benefits from its capacity to manage high-dimensional feature spaces through loss regularization. In the fine-tuned LLM case, it would be surprising for the ICL case to perform better, since it involves use the same models, with omission of the task specific training. Nevertheless, using ICL with general-purpose LLMs does not require any adaptation of the model or further training, proven to be a promising approach to quickly adapt LLMs to domain-specific problems. The literature supports our hypothesis that the efficacy of ICL likely stems from a nearest-neighbor-like mechanism \cite{Chen2023-wn, Moayedpour2024-fc}.

Because KRR does not produce uncertainty estimates, it is less suitable for BO, and we therefore do not explore it further. Additionally, due to the high output token cost associated with OpenAI fine-tuned models and our focus on ICL, we also do not employ the fine-tuned \texttt{gpt-3.5-turbo-0125} model for the BO task \cite{UnknownUnknown-gj}.

Testing on both the solubility and OCM datasets demonstrates that common machine learning performance metrics improve as the number of available few-shot examples increases (Figure ~\ref{fig:regression-results}). For example, using the OCM dataset, we observe improvements with newer OpenAI models. Specifically, \texttt{gpt-3.5-turbo-0125} achieves a mean absolute error (MAE) of $2.219\pm0.137$ and a correlation of $0.555\pm0.048$, whereas the newer \texttt{gpt-4o} attains an MAE of $1.863\pm0.151$ and a correlation of $0.649\pm0.060$ (see Table~\ref{tab:results} for complete results). 
Additionally, \texttt{gemini-2.5-flash} performs similarly with OpenAI models in the regression task, but shows better calibration, supported by the observed smaller negative log likelihood. This is an interesting characteristic for BO.
With the exception of krr, \texttt{gpt-4o} outperforms all other baselines in this study (see Figure~\ref{fig:regression-results} and Table~\ref{tab:results}). These results support our hypothesis that expanding the model's accessible memory pool (context)  thereby increases the probability of retrieving more query-relevant examples, and simulates a form of continual learning. This scaling capability is particularly important for BO. Although the retrieval-augmented ICL approach does not update the models' internal parameters over time as in traditional learning, ICL is a practical and effective strategy for adapting new data and overcoming the inherent constraint posed by an LLM’s fixed context window.

Our regression results indicate that LLMs can predict properties and directly produce uncertainty estimates from natural-language inputs. Additionally, in scenarios with abundant labeled data, ICL outperforms established methods such as Gaussian process regression (GPR) when applied to experimental procedure embeddings. Thus, we apply BO directly on language-based representations to maximize material properties within the OCM, AII, and RWGS datasets.

\subsection{Bayesian Optimization}\label{sec:bo}

We first apply BO-ICL to the OCM dataset, which provides a high-fidelity, unambiguous environment for initial evaluation, where in this setting, querying the "black-box" function $f(x)$ simply involves accessing the labeled dataset. Details regarding BO-ICL nomenclature and the algorithm are provided in Section~\ref{sec:methods}. Next, to address potential data leakage, we apply BO-ICL to optimize procedural parameters for two additional scenarios: A synthetic dataset representing alloy interface interactions (AII) and an in-house dataset focused on discovering optimal synthesis and reaction conditions to maximize $CO$ yield under RWGS reaction conditions.

\subsubsection{Oxidative Coupling of Methane}\label{sec:ocm_bo}

When testing on the OCM dataset, our goal in applying BO-ICL is to rediscover the optimal experimental conditions for maximizing the yield of value-added $C_2$ products (chemical equation \ref{eqn:oxi-meth_main}).

\begin{equation}
\text{(Oxidative Coupling of Methane)} \quad 2\mathrm{CH}_4 + \mathrm{O}_2 \rightarrow \mathrm{C}_2\mathrm{H}_4 + 2\mathrm{H}_2\mathrm{O} \quad 
\label{eqn:oxi-meth_main}
\end{equation}

\begin{figure}[h!]
    \centering
    \includegraphics[width=\linewidth]{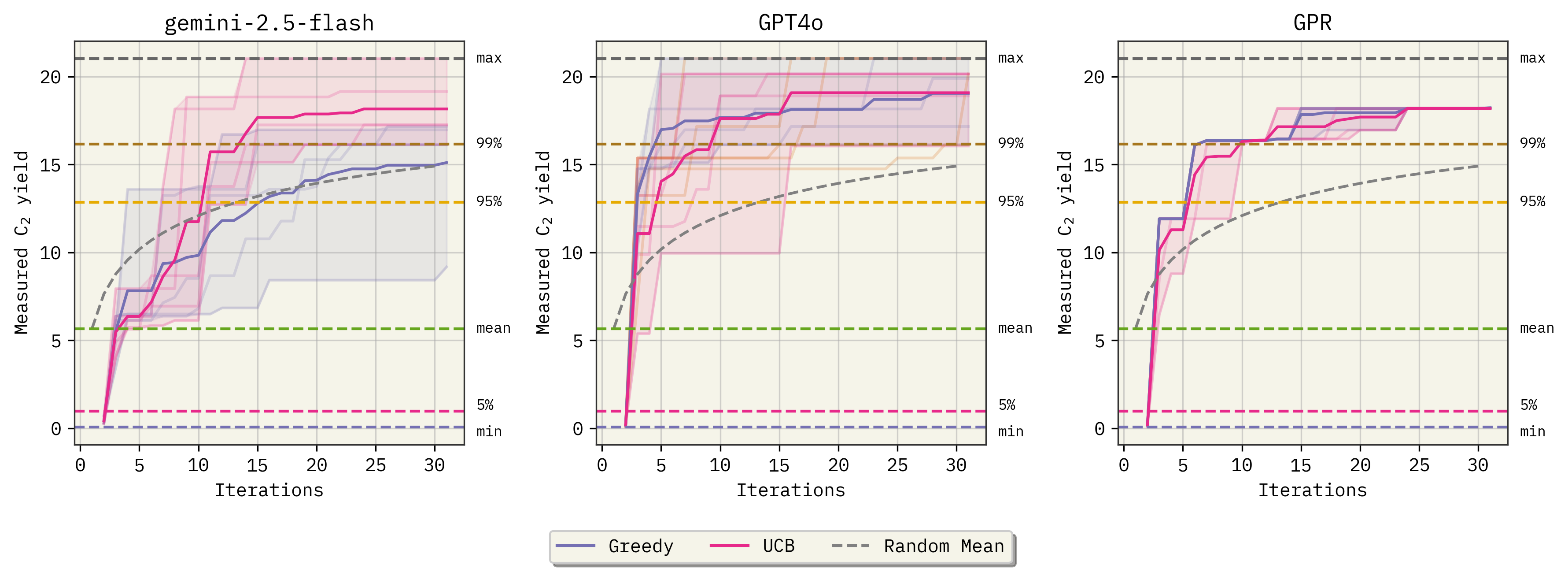}
    \caption{Bayesian optimization results for the OCM dataset. All results use an embedded natural language representation of the sampled experimental procedures as the input feature representation. We see convergence rates improve when using \texttt{gpt-4-0125-preview} instead of GPT3.5-turbo (data not shown).  While Gemini-2.5-flash requires, on average, 15 iterations to achieve the  $99^{th}$ percentile of the OCM dataset distribution, both GPT4o and GPR achieve this goal  after only 10 new samples, on average. Additionally, this figure implies that GPR using LLM embeddings performs satisfactorily (for GPR specifics see Section~\ref{ssec:gpr}).}
    \label{fig:bo-ocm}
\end{figure}

Thus, after converting the tabular \citet{Nguyen2020-kf} dataset to an unlabeled pool of possible experiments represented in natural language, we show that using an LLM as a surrogate model for BO is comparable to using GPR with identical feature vector representations. GPR is renown as a surrogate model for BO applications and thus is a reasonable baseline for performance analysis \cite{Li2024-vf}\cite{Wang2023-rl}\cite{Frazier2018-qa}. Results are shown in Figure~\ref{fig:bo-ocm}. Details about the dataset can be found in Section~\ref{sec:datasets}.

Applying BO-ICL to the OCM dataset demonstrates that \texttt{gpt-4-0125-preview} improves convergence rates toward higher $C_2$ yields over gemini-2.5-flash. This corroborates with our findings in the early regression experiments (Section ~\ref{sec:regression}). When using the upper confidence bound (UCB) acquisition function and iterating the BO loop for 30 new samples, Gemini-2.5-flash, on average, reaches the top $36^{th}$ procedure in the dataset, while GPT4o achieves the top $12^{th}$. These experimental procedure rankings correspond to a $C_2$ yield of 18.16 and 19.08, respectively. It is worth noting that even though \texttt{gpt-4-0125-preview} outperforms Gemini-2.5 on average, Gemini was able to find the top 1 procedure in one of the replicates. Comparatively, GPR's best selected point corresponds to a $C_2$ yield of 18.19 (top $33^{rd}$). On average, both the UCB and Greedy acquisition functions (Section ~\ref{sec:methods}) result in the same final procedure selection with either GPT4o or GPR surrogates. However, with \texttt{gpt-4-0125-preview}, the best possible procedure in the pool of approximately 12.8k examples is selected using the greedy acquisition function in three of the five replicates. 

These results imply that optimizing experimental procedures using language-based representation is a feasible method for optimizing experimental design. It is also evident that using embedding representations for GPR is also effective for property prediction and may offer the added advantage of reproducible results. However, LLMs may still be preferable over GPR for catalytic applications due to their ability to produce comparable results without requiring kernel tuning or other complex hyperparameter optimizations associated with GPR. Thus, BO-ICL is a straightforward and ready-to-use BO strategy for property prediction in complex material spaces.

Because the OCM dataset includes catalytic parameters that are well-established in literature, questions arise regarding the extent to which field biases may affect BO-ICL performance. In particular, prior catalysis studies on oxidative coupling of methane (OCM) often highlight $Mn$-$Na_2WO_4$ as a high-performing catalyst, with many OCM studies published before the GPT4o knowledge cutoff date \cite{Yunarti2017-bm, Sourav2023-kc, Nguyen2020-kf}. Notably, BO-ICL often converges on the $Mn$–$Na_2WO_4/SiO_2$ catalyst. This raises the question of whether the strong performance of BO-ICL could be attributed to data leakage. Although this seems unlikely given the transformation of tabular data into natural language, and the variance of catalytic performance across published results, we extend our workflow to the AII dataset. We expect the AII dataset to minimize the effects of leakage because the dataset is based on a less commonly used analytical equation to model interfacial material properties (Section~\ref{sec:aii_bo}).

\subsubsection{Estimated Alloy Interface Interaction}\label{sec:aii_bo}

Using a capacitor model to describe an alloy interface, as proposed by \citet{Gerber2023-xl}, we use BO-ICL to relate alloy-material pairs to the maximum unidirectional charge transfer in a pool of 9k alloys. The model approximates the calculated charge transfer labels using only Fermi levels, the transfer gap (defined as the sum of the largest van der Waals radii of the alloys), and the alloy stoichiometric chemical formulas, each specified in natural language (see Section~\ref{sec:datasets} for details).

\begin{figure}[h!]
    \centering
    \includegraphics[width=\linewidth]{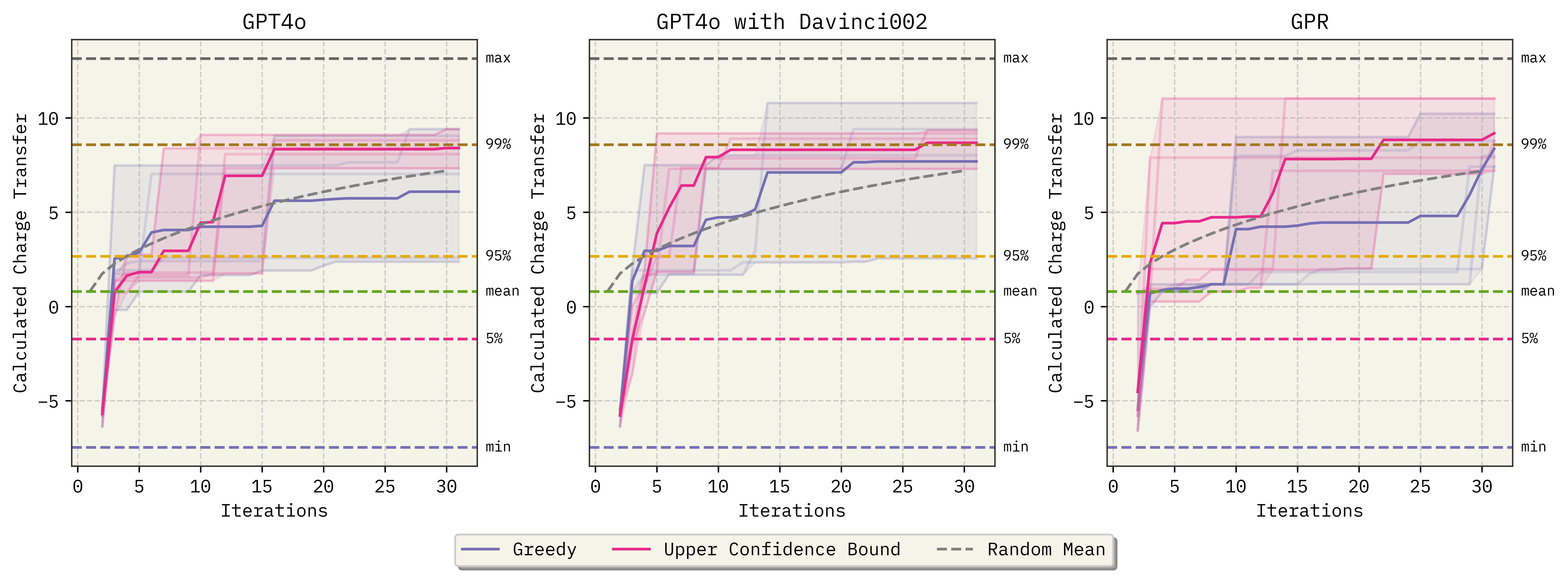}
    \caption{Results for the alloy interface charge transfer dataset (AII) using BO-ICL and GPR with natural language embeddings. We see comparable convergence rates and final property values selections within 30 BO iterations. Far left: Results using \texttt{gpt-4-0125-preview} at both the property value prediction step and inverse design step. Center: Results using \texttt{gpt-4-0125-preview} at inverse design step and \texttt{davinci-002} for property value completion with uncertainty estimations. Far right: Results using GPR (for GPR specifics see Section ~\ref{ssec:gpr})}
    \label{fig:bo-alloy}
\end{figure}

The AII dataset directly addresses concerns of data leakage, ensuring that performance improvements are not exclusively driven by strong field biases that are potentially encoded during pre-training. Since the original dataset is published after the \texttt{gpt-4-0125-preview} knowledge cutoff, the AII dataset is absent from the LLMs' pre-training data. Additionally, we incorporate alloy Fermi levels from the Materials Project database, as these values are not explicitly available in the original manuscript \cite{MaterialsProject2014,Jain2013}. Moreover, the analytical model for describing the alloy interface charge transfer relationship deliberately excludes spin-orbit coupling effects to simplify the model, which are known to influence band structure and charge transfer. This simplification, together with the logarithmic scaling of the charge transfer labels, reduces the potential for data leakage and bias from domain-specific training data. Therefore, the AII dataset is an appropriate application for evaluating the general efficacy of BO-ICL in less familiar knowledge domains. The rediscovery of material pairs within the top $99^{\mathrm{th}}$ percentile of the AII dataset underscores the robustness of BO-ICL in effectively guiding materials design.

Using the AII dataset, we also explore the use of LLMs better suited for different inference steps within the BO-ICL workflow (results are in Figure~\ref{fig:bo-alloy} - Center); in contrast to the exclusive use of \texttt{gpt-4-0125-preview} at each inference step when testing BO-ICL on all other datasets (Figures~\ref{fig:bo-ocm}) \cite{OpenAI2023-nv}. In this instance, for the property value prediction and uncertainty estimation step (flowchart step: \hyperref[fig:flowchart]{$A7$}), we use the \texttt{davinci-002} base model due to its superior calibration (i.e., the model’s predicted uncertainties closely align with actual prediction errors) in comparison with models like \texttt{gpt-4-0125-preview}, which are fine-tuned using reinforcement learning from human feedback (RLHF) \cite{OpenAI2023-nv}. RLHF can introduce biases that prioritize human-aligned responses over strictly probabilistic accuracy, potentially degrading a model’s ability to produce well-calibrated uncertainty estimates \cite{OpenAI2023-nv}. Our decision to incorporate \texttt{davinci-002} comes from the observed importance of model calibration on overall performance (see Section~\ref{sec:methods} and SI). Using a well-calibrated off-the-shelf model for the regression step alleviates the need for post-training calibration and reduces the number of initially labeled data points required to achieve satisfactory performance. For the inverse design generation step (flowchart step: \hyperref[fig:flowchart]{$O1$}), we continue the use of \texttt{gpt-4-0125-preview}, as its RLHF training ensures an output structure that more closely aligns with the natural language format of the experimental procedures. This alignment is particularly useful for the similarity comparison and retrieval steps in the optimization loop (Figure~\hyperref[fig:flowchart]{$A2-O3$}, Algorithms~\ref{alg:mmr},~\ref{alg:create-sub-pool}). The performance differences when using a single model (\texttt{gpt-4-0125-preview}) versus a combination of a base model and a chat model (\texttt{davinci-002} and \texttt{gpt-4-0125-preview}) in the workflow may further highlight the critical role of accurate uncertainty estimations when comparing upper confidence bound (UCB) trajectories (see Section ~\ref{sec:methods} for acquisition function details). It is important to note that observed performance on a dataset like AII may relate to the use of a well-defined analytical objective function, as opposed to the other datasets relying on experimental labels, which are more susceptible aleatoric measurement errors. Although direct comparison between the use of different datasets and models remains challenging due to replicate limitations and inherent model differences, achieving performance that outpaces random-walk baselines on complex datasets like AII is sufficient motivation to synthesize and test materials in-house, using BO-ICL to guide the experimental parameter selection for optimizing catalyst synthesis and reaction conditions (Section ~\ref{sec:rwgs_bo}).

\subsubsection{In-house RWGS}\label{sec:rwgs_bo}

To extend our workflow to scenarios where experimental outcomes are unknown a priori, we apply BO-ICL to a pool of experiments where we synthesize and test catalysts on demand for the RWGS reaction (chemical equation). 

\begin{equation}
\text{(Reverse Water-Gas Shift)} \quad \text{CO}_2 + \text{H}_2 \rightleftharpoons \text{CO} + \text{H}_2\text{O} \quad 
\label{eqn:RWGS}
\end{equation}

Our objective is to maximize $CO$ yield, the desired product from RWGS. Since the LLM has no prior exposure to this particular experimental space, the model's performance primarily reflects the optimization policy's ability to leverage GPT's general knowledge of catalysis. Additionally, these experimental settings provide insight into how well BO-ICL accounts for human experimental error when selecting the next experiment from the pool. Further details on the experimental setup are provided in the SI (Section~\ref{sec:datasets})

\begin{figure}[hbt!]
    \centering
    \includegraphics[width=.65\linewidth]{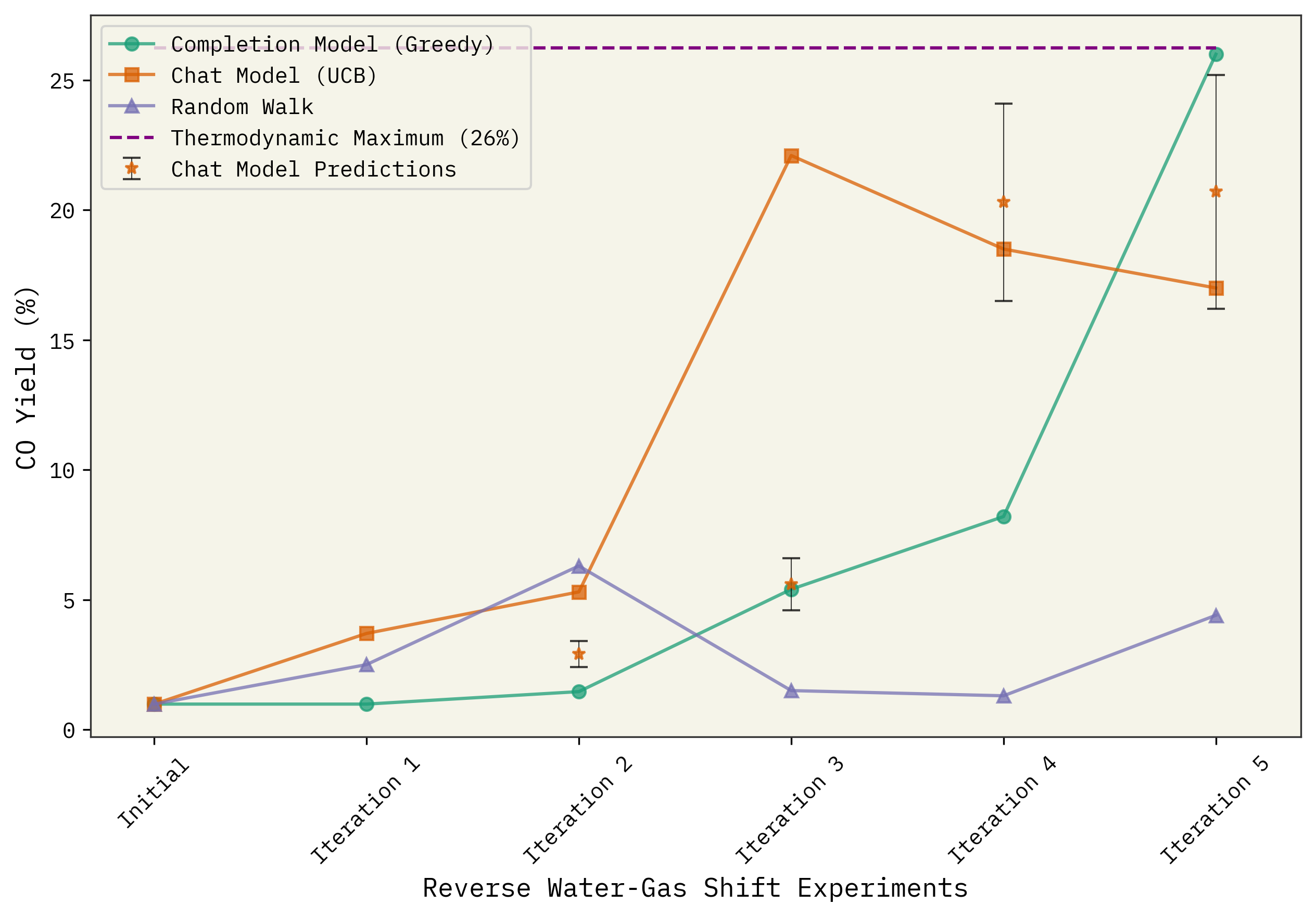} 
    \caption{BO-ICL results on an unlabeled pool of experimental combinations for the reverse water-gas shift (RWGS). We present results from three independent runs. In purple, 6 experiments are randomly selected using a Python's built in Mersenne Twister pseudorandom number generator. In green, we show the trajectory using the now deprecated model \texttt{gpt-4-32k-0314}, with the greedy acquisition function. A similar analysis using \texttt{gpt-4-0125-preview} is displayed in orange. \texttt{gpt-4-0125-preview} $C_2$ yield predictions with uncertainty estimations for each sample, prior to carrying out the experiment, are displayed as orange stars and error bars for the EI acquistion function. The dashed horizontal line represents the thermodynamic maximum CO yield possible under the RWGS conditions used in this study.}
    \label{fig:in_house} 
\end{figure}

Figure~\ref{fig:in_house} presents three trajectories: a random walk (purple), BO-ICL using the base model (\texttt{gpt-4-32k-0314}- green), and a chat model (\texttt{gpt-4-0125-preview}-orange). The random walk represents a series of experiments that are chosen using a random number generator to provide some insight into the sample space distribution. We apply BO-ICL with the now-deprecated \texttt{gpt-4-32k-0314} model and observe monotonic performance when using the Greedy acquisition function. This observation is consistent with Greedy's design, which ignores prediction probabilities and explicitly optimizes through exploitation (Section ~\ref{sec:methods}). We also run BO-ICL using the same sample pool with the later released \texttt{gpt-4-0125-preview} model to ensure reproducible performance with use of chat models. Pairing this chat model with the Upper Confidence Bound (UCB) acquisition function suggests that exploration is a priority in procedure selection (iterations 5 and 6). Since UCB incorporates uncertainty estimations as a parameter (Section ~\ref{sec:methods}), we include the model's original mean prediction value (star) and the corresponding uncertainty estimate (error bars) for each procedure selection. 

It is important to note that each experiment appears in the order selected by BO-ICL. Both models demonstrate CO yields over 20\% within six experiments, approaching the maximum thermodynamically achievable CO yield under these conditions (Table ~\ref{tab:inhouse-results}). These results strongly support the potential efficacy of BO-ICL in real-world applications.

\section{Limitations} 

BO-ICL inherits limitations intrinsic to its sub-components: principled BO and the nondeterministic behavior of LLMs. For instance, a key approach in BO frameworks is ensuring sufficient diversity in the initial dataset ~\cite{Konakovic-Lukovic2020-hb, Morishita2023-ls}. While ICL typically benefits from similarity among contextual prompt examples, early implementations of BO-ICL indicate that insufficient diversity within the initially available context pool severely constrains exploration, despite adjustments in acquisition function parameters aimed at enhancing exploration (e.g., $\lambda$ parameter in UCB; see section \ref{sec:methods}). Consequently, this attribute may limit the full utilization of available data. BO-ICL requires that the initial available context pool is balanced, to adequately prevent local optimization. To address this, users can strategically adjust the BO-ICL initialization to begin with $N$ diverse labeled data points. Specifically, one may begin by populating the BO-ICL memory by first selecting a random data point from the available labeled data to use as a reference using \texttt{MMR} with $\lambda=0$, in attempt to match the distribution of diversity in the unlabeled pool(Equation ~\ref{eq:MMR}). In each BO experiment, we use this approach with $N=2$. Thus, in cases with significant amounts of initial labeled data, careful consideration of balancing data usage to fine-tune a model, thereby developing a domain-specific model, and initializing the context pool with available data is advised.

Additionally, for effective exploration for global optimization, accurate confidence estimates from surrogate models are essential in BO. However, as aforementioned, SOTA LLMs subject to RLHF frequently exhibit calibration challenges, complicating the generation of accurate uncertainty estimates, without an extensive validation data. Such a requirement conflicts with the primary advantage of BO, which is optimizing objectives with minimal data. We are able to gain some control over this concern by (1) employing models (e.g., \text{davinci-002}) that provide better calibrated uncertainty estimates leading to comparable performance with calibrated models without the additional calibration steps (relevant results visible in Section ~\ref{sec:aii_bo}), and (2) leveraging the transfer learning capabilities of LLMs when feasible. To re-illustrate method (2), as seen in Section \ref{sec:methods}, derivation of a calibration scaling factor was possible using the Uncertainty Toolbox, by sampling a validation set exclusively from the OCM dataset. This scaling factor is consistently useful in observing positive performance in BO-ICL applications beyond just the OCM dataset and even OpenAI models (i.e. we use this same scaling factor for every application in this investigation unless otherwise specified). This is interesting considering that calibration processes are commonly renown as subjective. We show that using available datasets similar to the design space of focus can be a valid method to overcome the call for initial data to achieve satisfactory calibrated models useful in the BO workflow.

Regarding LLM-specific drawbacks, hallucination remains a prominent issue. Hallucination affects are significantly visible at the inverse design step of BO-ICL in early applications (steps A1 - O1 in Figure ~\ref{fig:flowchart}). Hallucinations frequently result in irrelevant or infeasible completions, diminishing search efficiency for sub-pool population (visible in workflow steps O1-O2, Figure ~\ref{fig:flowchart} ) and scientific validity. To hedge hallucination influence, we may restrict inverse design outputs to predefined design parameters using custom system messages for the design space of focus (see SI Section ~\ref{sec:sysmes}). This significantly enhances the probability of a model to complete queries with procedures that correspond to an actionable experimental parameter combination within the sample space. This is especially important for the RAG action in BO-ICL (see process steps A4 - A8 in Figure ~\ref{fig:flowchart}). Thus, the added step of optimizing a system message or a method to control inverse designs may be necessary for effective implementation.

Evaluation challenges represent another relevant concern when employing LLMs due to their inherent stochasticity during token generation processes. Hyperparameters such as temperature and the chosen sampling strategy (e.g. topk) can help control variation, but cannot remove it. Thus, one to one reproducibility is unlikely, especially in use cases with closed source models. Surrogates like Gaussian-processes offer this attribute. For this reason, we empirically attempt to estimate average performance by running 5 replicates of every acquisition function. Although 5 runs may be too few for the central limit theorem to guarantee a normal sampling distribution, we are able to compute point estimates to carry out hypothesis testing using non-parametric methods, keeping in mind the limited statistical data. It may be important to highlight that the inherent randomness of LLM outputs can also positively influence exploration and novel discovery.

The notable strength, but also a key concern, of using these general purpose models as surrogates is their extensive base knowledge acquired through large scale pre-training. While this characteristic can be advantageous for rapid optimization in BO-ICL, it also introduces the risk of adapting domain-specific biases. Specifically, pretrained models may overcompensate during completion with favoring field-familiar material designs, thereby constraining exploration. At the inverse design stage, the model may preferentially suggest materials that are well-studied or prominent in existing literature, potentially limiting discovery to already established domains. The working hypothesis for continuing to use these models in novel material discovery is that exploratory acquisition functions, such as UCB, can help counteract such biases over multiple optimization iterations. Furthermore, prior studies have shown that when queried with examples less similar to their training data, LLMs increasingly rely on the provided context examples over base knowledge to improve completion accuracy \cite{Chen2023-wn, Du2024-xv}. A potentially effective strategy enhance more novel design combinations is careful curation of the design space by deliberately including less-studied material combinations, users can guide the model toward exploring novel regions of the design landscape.

However, this strategy underscores another important limitation, which is the necessity of strategic control over user ignorance or biases for design space construction. Poorly informed sampling of the design space, especially safety sensitive domains like catalysis, can lead to wasted resources or even hazardous conditions. For instance, when attempting to optimize an exothermic reaction, choosing reaction conditions or equipment parameters without considering thermal runaway risks or lacking consideration of material decomposition temperatures can lead to adverse outcomes. To mitigate these risks, it is strongly recommended to use traditional scientific methods, such as accumulation of prior data on material properties or extending collaboration for expert consultation and analysis, before curating and testing the design space. This concern is a standard in traversing any new design landscape.

\section{Conclusion}

This work introduces BO-ICL, a framework that integrates Bayesian Optimization (BO) with In-Context Learning (ICL) using large language models (LLMs) to optimize experimental conditions directly from natural language representations. We demonstrate the effectiveness of BO-ICL across four datasets: solubility (ESOL), oxidative coupling of methane (OCM), alloy interface interaction (AII), and reverse water-gas shift (RWGS). On the OCM dataset, BO-ICL reaches the 99th percentile of candidate procedures using only ten additional samples, matching the performance of Gaussian Process Regression (GPR) with natural language embeddings. Moreover, BO-ICL successfully guided real-world RWGS catalyst experiments, achieving CO yields near the thermodynamic limit after only six iterative experiments.

Our results highlight that LLMs are practical surrogates for BO by leveraging their scalability through example-based reasoning. Unlike traditional approaches, BO-ICL operates without feature engineering, architectural tuning, or retraining, making it a zero-shot and task-agnostic solution for design optimization in materials science. BO-ICL is a reliable and accessible framework for accelerating experimental design, using natural language as a universal chemical representation, enabling optimization with minimal computational resources, thereby eliminating the need for task-specific fine-tuning or feature selection. The framework is available open-source at https://github.com/ur-whitelab/BO-ICL.

\section{Methods}\label{sec:methods}

\subsection{Bayesian Optimization}

BO is a sequential, gradient-free strategy for optimizing an expensive to evaluate black-box function $f(x)$\cite{Frazier2018-hq}. BO is particularly useful in settings where direct evaluation of the objective function is costly, such as catalysis focused wet-lab research. 
BO aims to solve the optimization problem

\begin{equation}
\arg\max_{x \in \Omega} f(x)
\end{equation}

where $\Omega$ is typically a hyper-rectangle domain that limits the set of possible experiments. We call $\Omega$ the sample space.

In order to run BO, a surrogate model $\mathcal{S}(x)$ is used to approximate the expensive-to-evaluate black-box function $f(x)$. 
Surrogate models are often probabilistic, offering query predictions along with corresponding uncertainty estimations at inference. GP models are commonly used as surrogates (See Section~\ref{sec:baselines}).

Initially, the prior $\mathcal{S}(x)$ is trained using all already available data $\mathcal{D}$. Then the posterior probability distribution can be computed as $\mathcal{S}(x|\mathcal{D})$. On each iteration, the probabilistic model is used to compute a set of posterior probability distributions and an acquisition function $\alpha(x)$ is used to rank and select the next sample to evaluate. Most acquisition functions use the prediction mean ($\mu(x)$) and uncertainty ($\sigma(x)$) to balance the trade-off between exploring uncertain regions of the input space and regions where the surrogate model predicts high values for $f(x)$.

In this work, we focus on three acquisition  functions:
The Upper Confidence Bound (UCB), which balances exploration and exploitation by incorporating both the mean and uncertainty: $\alpha_{\mathrm{UCB}}(x) = \mu(x) + \lambda \sigma(x)$, where $\lambda$ is a tunable parameter that controls the exploration-exploitation trade-off.
Another acquisition function considered was the greedy acquisition function. This function always selects the point with the highest predicted mean from the surrogate model, favoring exploitation.  The greedy acquisition function can be expressed as: $\alpha_{\mathrm{greedy}} = \mu(x)$.
Lastly, we employed a random sampling as a baseline. The random sampling selects the next point to evaluate using a random number generator, to define an index to select from the sample space $\Omega$. In this case, the next experiment is selected as: $x_{\mathrm{next}} \sim \operatorname{Uniform}(\Omega)$

In the sequence, the black-box function $f(x)$ is evaluated to obtain the label for the selected point, which is then added to the training dataset $\mathcal{D}$ for the next iteration of the BO policy.

The BO algorithm proceeds iteratively as follows:

\begin{algorithm}[H]
\caption{Bayesian Optimization Policy for Reaction Runs}
\begin{algorithmic}[2]
\State \textbf{Input}: Initial dataset $\mathcal{D}$ \Comment{Initialized with two labeled points}
\Repeat
    \State $\mathcal{S} \leftarrow \mathrm{train}(\mathcal{D})$ \Comment{Update context for surrogate model}
    \State $x \leftarrow \operatorname*{arg\,max} \alpha(x; \mathcal{S}, \mathcal{D})$ \Comment{Select next reaction condition using acquisition function}
    \State $y \leftarrow f(x)$ \Comment{Run reaction and observe property value}
    \State $\mathcal{D} \leftarrow \mathcal{D} \cup \{(x, y)\}$ \Comment{Update labeled context}
\Until{termination condition is reached}
\State \textbf{return} $\operatorname*{arg\,max}_{\mathcal{D}} \mathcal{S}(x)$
\end{algorithmic}
\end{algorithm}

\subsection{BO-ICL workflow}\label{sec:boworkflow}

BO-ICL leverages LLMs as surrogate models for BO of select parameters. We use ICL to dynamically update the posterior at inference using labeled examples. To ensure scalability with new data, we implement a long-term memory of labeled samples, allowing the use of relevant context for prompt construction. By dynamically generating prompts, we show that model performance can improve even beyond its context window (i.e., the maximum amount of input data the model can process at once) as new data is acquired (Section ~\ref{sec:regression}). 

We use cosine similarity with the query of focus as the reference to down-sample the labeled pool for prompt generations. Thus, for each query, often an unlabeled experimental procedure, we identify the most relevant examples and prefix them for ICL at inference time. This prompt generation process uses LangChain \cite{Chase_LangChain_2022} and the available FAISS library \cite{johnson2019billion}, along with Ada-002 embeddings \cite{neelakantan2022text}.

The queries follow a general prompt structure for LLM input: \texttt{\{prefix\}\{few-shot template\}\{suffix\}}. The \texttt{\{prefix\}} provides instructions and constraints for the task, including the expected response format, to minimize hallucinations. This step, often implemented as a \texttt{system\_message}, is especially important for guiding chat model behavior. Including the task description in the \texttt{system\_message} significantly improves performance ~\ref{sec:sysmes}. 

The \texttt{\{few-shot template\}} formats the context by concatenating $k$ examples using the following structure: ``\texttt{Given \{representation\}. What is \{property\_name\}? \{completion\}}''. Figure~\ref{fig:toc} illustrates how the prompt is constructed by selecting $k = 1$ examples as context. Finally, the \texttt{\{suffix\}} contains the primary query of interest for which the LLM should provide a completion.

For the regression steps with uncertainty, we use token probabilities, following an approach similar to the action selection process described in \citet{ahn2022can}. To estimate model uncertainty, we marginalize the log probabilities of the completion tokens to derive a discrete probability distribution after $n$ iterations(Equation ~\ref{eqn:wdev}). This distribution can then be leveraged for weighted uncertainty approximations, which are directly applied within the acquisition functions for BO \cite{Frazier2018-hq}.
\begin{equation}\label{eqn:wdev}
\sigma = \sqrt{\frac{\sum_{i=1}^{N} w_i (x_i - \bar{x}^*)^2}{\frac{(N-1)}{N} \sum_{i=1}^{N} w_i}}
\end{equation}

where  $N$ is Total number of observations, $x_i$ means the value of the $i$-th observation. We represent the weighted mean of the observation as $\bar{x}^*$, calculated as \( \bar{x}^* = \frac{\sum_{i=1}^{N} w_i x_i}{\sum_{i=1}^{N} w_i} \). Finally, $w_i$ is the weight assigned to the $i$-th observation, reflecting its relative importance or observation probability. 

Finally, these methods are combined into a BO loop to optimize experimental parameters. This is advantageous because the BO approach requires no traditional training and has minimal compute requirements for inference. A flow chart illustrating the implementation of BO-ICL is provided in Figure~\ref{fig:flowchart}, and a pseudo-code implementation is available in Algorithm~\ref{alg:bo-icl}.

BO-ICL starts by using an optional labeled dataset $\mathcal{L}$ to populate the LLM long-term memory $\mathcal{M}$ (step $A1$ in Figure~\ref{fig:flowchart}). If $\mathcal{L}$ is not available, the LLM initiates the optimization without prior knowledge of the space of possible experiments. Typically, the surrogate model is used to evaluate the entire space of possible examples $\mathcal{U}$. However, due to the computational cost of using LLMs and the latency associated with API calls, we adopt an embedding-similarity retrieval approach to sub-sample $\mathcal{U}$ for the regression step (steps $A5$–$A7$).

We create a sub-pool by using MMR, with an inverse-designed completion serving as the reference embedding for retrieval \cite{murray2005extractive, Guo2010-zt}. MMR aims to reduce redundancy in the sampled set while ensuring the selected points remain relevant to the query. We use cosine similarity to compare the Ada embedding representations. 
MMR is computed as shown in Equation~\ref{eq:MMR}, and a pseudo-code implementation is provided in Algorithm~\ref{alg:mmr}.

\begin{equation}\label{eq:MMR}
    \mathrm{MMR} = \operatorname*{arg\,max}_{d_i \in \Omega \setminus S} \left[ \lambda \cdot \mathrm{Sim}(d_i, q) - (1 - \lambda) \cdot \max_{d_j \in S} \mathrm{Sim}(d_i, d_j) \right]
\end{equation}

To obtain this reference procedure, we first search $\mathcal{M}$ for examples with labels similar to the current best label $y^+$ (step $A2$). These examples are used as context to query a new procedure $x'$ corresponding to a slightly higher predicted label $y'$, defined as:

\begin{equation}
y' = y^+ + \left(|y^+|\cdot \mathcal{N}(0.2, 0.05)\right), \quad x' = \mathrm{LLM}(y' \mid \mathcal{M})
\end{equation}

Here, $x'$ is the inverse-designed input (object $O1$), representing a hypothesized experiment with a label greater than $y^+$. We then use $x'$ as a reference to retrieve $n$ similar experiments from $\mathcal{U}$ using MMR (steps $A4$ and $A5$). These $n$ experiments form the sub-pool (object $O2$), which is passed to the regression step (step $A7$) to select the next experiment (object $O3$). As with the inverse design step, we construct a dynamic prompt context for each experiment $x$ in the sub-pool by searching $\mathcal{M}$ for the most similar examples (step $A6$), using cosine similarity. The LLM is then used to predict a label $y$ for each $x$ in the sub-pool (step $A7$), and these predictions are scored using an acquisition function $\alpha$. The top $n$ candidates, based on $\alpha$, are selected (step $A8$).

\begin{center}
  \includegraphics[width=\linewidth]{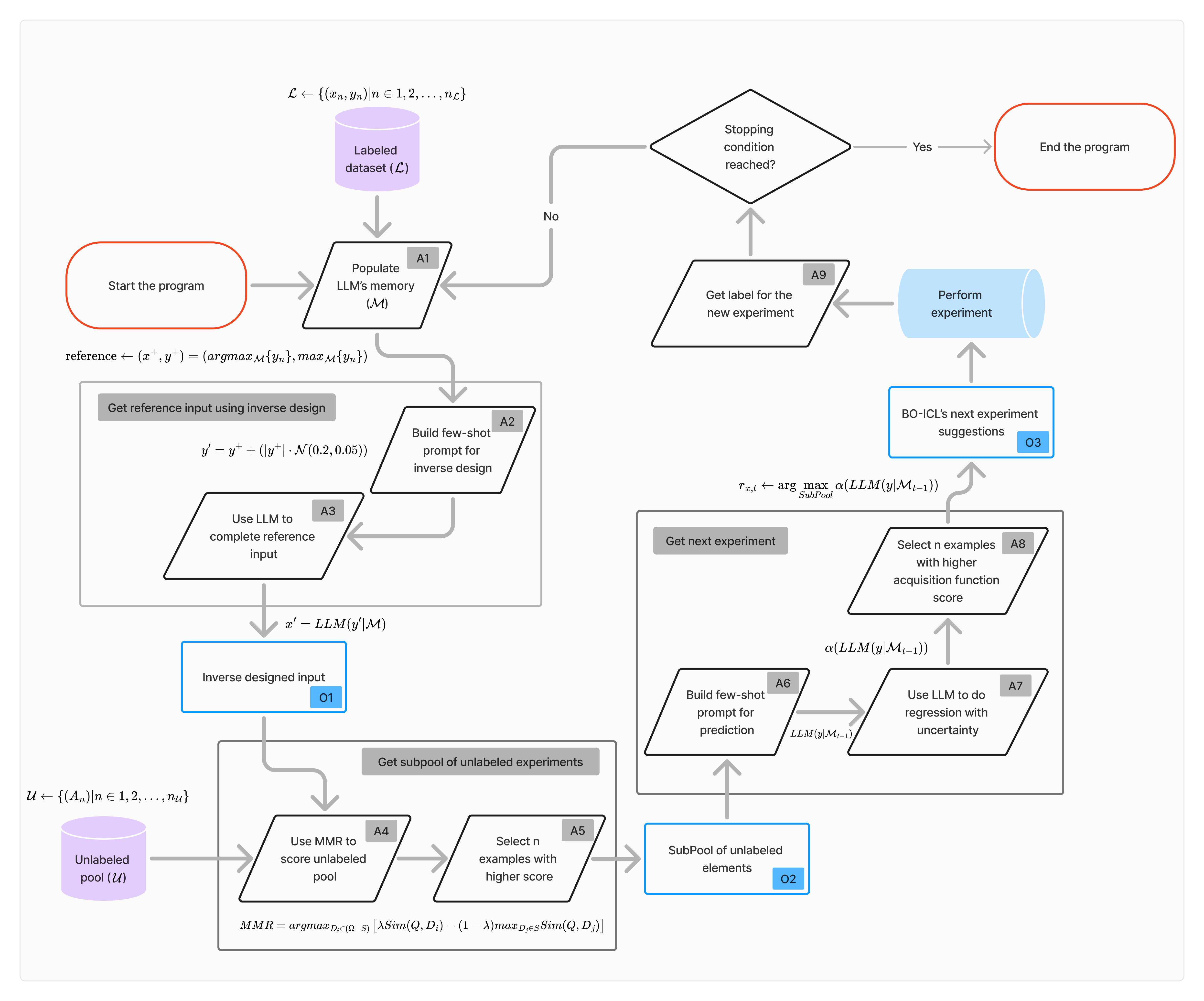}
  \captionof{figure}{Flow chart diagram of the information flow in BO-ICL. Angled black rectangles represent actions, blue rectangles highlight key objects, used in the workflow. Some actions with common goal are grouped together within a gray box, with a label describing its goal. Actions are identified using the $An$ indexer, and objects use a $On$ syntax. $n$ is an index without further meaning. The same pipeline is shown in Algorithms~\ref{alg:llm}, \ref{alg:create-sub-pool}, and \ref{alg:bo-icl} using pseudo-code.}
  \label{fig:flowchart}
\end{center}

Next, we obtain the ground-truth labels for the selected experiments (step $A9$). For the ESOL, OCM, and AII datasets (see Section~\ref{sec:datasets}), the label is directly queried from the available datasets. In the case of the in-house RWGS unlabeled dataset, the experiments proposed by BO-ICL are physically run and analyzed to determine the corresponding labels (step $A9$). The optimization loop continues until a specified stopping criterion is met (e.g., when the sample selected maps to the maximum possible performance in the system). Until that point, newly labeled experiments are added to $\mathcal{M}$, and the loop proceeds. Upon reaching the stopping condition, the experiment with the highest observed label $y^+$ is retrieved from $\mathcal{M}$.

\subsection{Hyperparameter tuning}\label{sec:hyperparam}
Our algorithm requires defining key hyperparameters, including the number of few-shot examples ($k$) used as context and the temperature ($T$), which controls sampling for the LLM's output. To investigate the effects of these hyperparameters, we conducted a systematic study by varying both $k$ and $T$ using \texttt{gpt-3.5-turbo-0125}, given its reduced cost.

\begin{center}
  \includegraphics[width=0.8\linewidth]{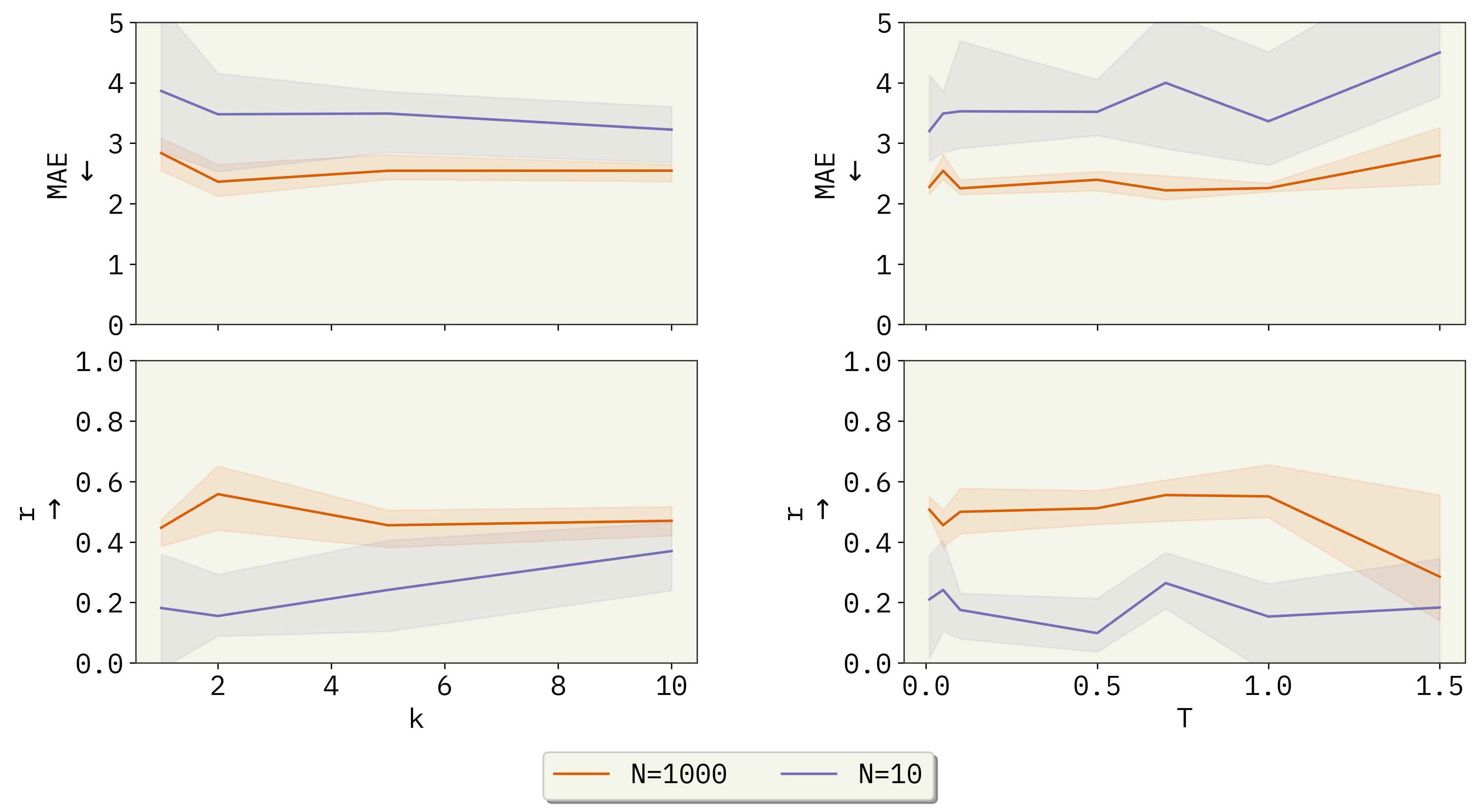}
  \captionof{figure}{Analysis for hyperparameter selection. This figure shows
    the results of varying $k$ (the number of context examples per prompt) and
    the temperature ($T$) for \texttt{gpt-3.5-turbo-0125}, which controls the
    spread of the output distribution over the model’s vocabulary to tune the
    degree of randomness.}
  \label{fig:ocm-metrics}
\end{center}

For the systematic study, we first fixed $T = 0.05$ and $N = 1000$ for the OCM dataset, or $N = 700$ for ESOL. The orange curves in Figure~\ref{fig:ocm-metrics} show that our system is weakly influenced as a function of $k$. Results for both $k = 5$ and $k = 10$ lie around a mean absolute error (MAE) of $\sim 2.5$ and a correlation of $\sim 0.5$. These two results are not statistically different, with a p-value of $0.985$ (Table~\ref{tab:pvalue}).

This result is somewhat counterintuitive. To further investigate why the number of examples in context does not affect the model, we performed the same analysis but added only ten random examples to the LLM's memory. Figure~\ref{fig:ocm-metrics} (blue curve) shows a small MAE decrease from $3.490 \pm 0.380$ to $3.224 \pm 0.361$, while the correlation increased from $0.241 \pm 0.114$ to $0.370 \pm 0.073$, likely highlighting the importance of context in the low-data regime. These results corroborate with literature in observation of diminishing returns from extended context lengths \cite{Baek2024-hw}.

These results, along with the relationships shown in Figure~\ref{fig:regression-results}, may indicate varying degrees of bias influenced by the model’s pre-training familiarity with different datasets. For example, the solubility dataset, where correlation values for GPT-4o reach $0.9$ (Figure~\ref{fig:sol-metrics}) with minimal available examples, suggests a higher level of familiarity compared to OCM (where $r \approx 0.6$). This aligns with the expectation that models rely more on prior knowledge in familiar settings, but depend more heavily on in-context data in less familiar test spaces \cite{Du2024-xv}.

Similarly, we fixed $k = 5$ to run the systematic study for $T$. The T-test studies (Table~\ref{tab:pvalue}) show that differences in results for experiments with $T$ within the range $0.1$ to $1.0$ are not statistically significant. However, we observed a considerable decrease in performance for $T > 1.0$ (Figure~\ref{fig:ocm-metrics}), caused by increased hallucination in the LLM outputs. The temperature variation effects are also related to the degree of model calibration.

We acknowledge that some of the models explored in this study were trained using reinforcement learning from human feedback (RLHF), which can lead to less calibrated probability estimates during inference \cite{OpenAI2023-nv, Kapoor2024-iw}. Instruction tuning with RLHF may introduce biases in a model's output probability distribution due to subjective human annotations, potentially resulting in poor confidence estimates \cite{OpenAI2023-nv}. Given that BO policies rely on accurate likelihood representations, we first sought to quantify the calibration of relevant models, using uncertainty estimations extracted as mentioned in Section ~\ref{sec:boworkflow}.

To assess the level of miscalibration between the predictive methods for uncertainty extraction, we utilized the `Uncertainty Toolbox' (UCT)\cite{Kuleshov2018-yi, Chung2021-pa} package. UCT provides tools to calculate calibration metrics such as calibration error and prediction interval coverage probability. Validation samples were grouped based on their model prediction uncertainties to form confidence intervals for binning inferred values. The model's prediction accuracy was then evaluated for samples that fall within each confidence interval to analyze how well the predicted intervals align with observed outcomes. The relationship between the predicted and observed proportions was used to plot the calibration curve and to compute the miscalibration area (MA), which quantifies the deviation from the ideal, monotonic calibration curve.

\begin{figure}[H]
    \centering
    \includegraphics[width=\linewidth]{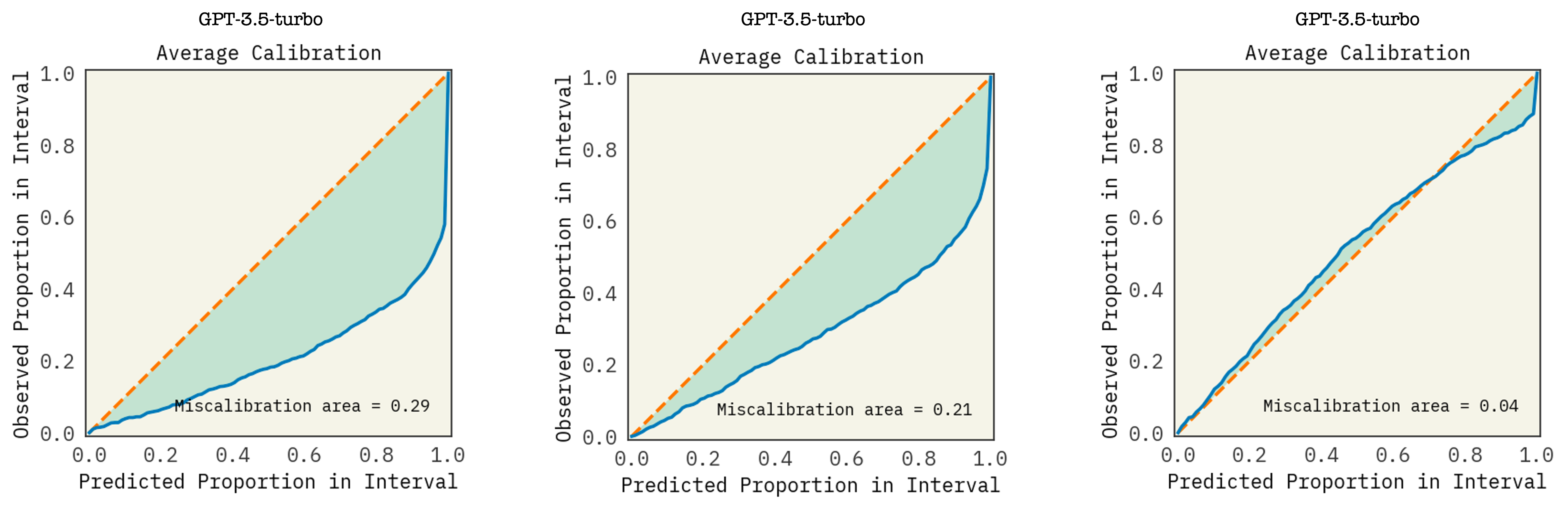}
    \caption{This shows the differences in calibration levels for GPT-3.5-turbo using 1k points from the OCM dataset, using five predictions for each prompt. Here, calibration evaluation involves three methods: (left) conditional probabilities to calculate weighted standard deviations and represent a confidence score for a single prediction; (center) standard deviation calculated using consistency across the predictions; and (right) optimizing for a scaling factor and applying it to assess the degree of calibration using a calibration dataset of 25 examples. Beyond the 25 samples for calibration, no significant improvement was observed, so the use of 25 points for calibration was continued.}
    \label{fig:calibration}
\end{figure}

The MA can then guide the optimization of an uncertainty scaling factor, expected to enhance calibration. Figure ~\ref{fig:calibration} illustrates calibration differences with and without applying this scaling factor, using 1000 points from the OCM dataset for evaluation, along with a comparison of the uncertainties using the two aforementioned extraction methods. A validation set (25 samples) used from the OCM dataset was used to optimize this scaling factor; beyond the use of 25 points exhibited nominal variation in the MA of GPT-3.5-turbo. Interestingly, applying this calibration factor during testing of BO-ICL across different datasets consistently displayed performance improvements (add evidence to the SI). This observation is notable, as calibration is often considered a subjective process, with parameter effectiveness typically varying between tasks and datasets. The ability to calibrate models effectively using a small number of samples from a single dataset, may further indicate the transfer learning potential of these SOTA LLMs.

As supported in the literature, using simple consistency arguably offers a greater degree of calibrated uncertainties over a model's inferred distribution \( p(y_i \mid \theta, x_i) \) following preference or instruction tuning (Figure~\ref{fig:calibration}) \cite{Farquhar2024-mb, QingLyu-oe}. 

Based on this analysis, we defined the hyperparameters as \( k = 5 \), \( T = 0.7 \), and a calibration factor of 5. These values were used for all BO experiments presented in the main paper.

%% file: supMat.tex
\SIsection{BO-ICL algorithms}\label{sec:icl_algs}

\begin{algorithm}
    \caption{Algorithm to compute MMR. CosSim is cosine similarity.}
    \label{alg:mmr}
    \begin{algorithmic}
        \State \textbf{Initalize:} 
            \State \hspace{1cm} penalty $\gets$ 0
        \State \textbf{Input:} 
            \State \hspace{1cm} Element for which MMR is being computed: $e$
            \State \hspace{1cm} Reference element: $r$ 
            \State \hspace{1cm} Already selected elements: $S$
        \Function{ComputeMMR}{$e$, $r$, $S$}
        
            \For{s in S}
                \State curr\_penalty $\gets$ CosSim($r$, $s$)
                \If{curr\_penalty > penalty}
                    \State penalty $\gets$ curr\_penalty
                \EndIf
        
            \EndFor
            \State mmr\_score $\gets$ $\lambda$CosSim($r$, $e$) - $(1-\lambda)$penalty
            \State \textbf{return} mmr\_score
        \EndFunction
    \end{algorithmic}
\end{algorithm}

\begin{algorithm}
    \caption{LLM prediction algorithm}\label{alg:llm}
    \label{alg:LLM}
    \begin{algorithmic}
        \State \textbf{Initialize:} 
            \State \hspace{1cm} context examples: K $\gets$ []
            \State \hspace{1cm} LLM provider: api
            \State \hspace{1cm} Task description: prefix
        \State \textbf{Input:} 
            \State \hspace{1cm} LLM's memory:: \(\mathcal{M} \gets \{(\eta_x, \eta_y)_n | n \in {1,2,...,n_{\mathcal{M}}}\}\)
            \State \hspace{1cm} context size: $k$
            \State \hspace{1cm} Element for which the prediction needs to be done: $input$
        \Function{LLM}{$input$, $k$, $\mathcal{M}$}
            \LComment{Create context for ICL}
            \While {length(K) < k}
                \State new\_example $\gets$ $\arg\max_L$ ComputeMMR($\mathcal{M}$, input, K)
                \State K.append(new\_example)
            \EndWhile
            \\
            \LComment{Prepare the propmt}
            \State query $\gets$ prefix + examples(K) + suffix(input)
            \\
            \LComment{Send a request to the LLM provider}
            \State completion $\gets$ api.request(query)
            \State \textbf{return} completion 
        \EndFunction
    \end{algorithmic}
\end{algorithm}

\begin{algorithm}[H]
    \caption{Sub-Pool creation algorithm}\label{alg:create-sub-pool}
    \begin{algorithmic}
            \State \textbf{Initialize:} 
            \State \hspace{1cm} counter: k $\gets$ 0
            \State \hspace{1cm} SubPool $\gets$ [] \Comment{Empty list}
        \State \textbf{Input:} 
            \State \hspace{1cm} Pool of unlabaled data: \(\mathcal{U} \gets \{(A_n) | n \in {1,2,...,n_U}\}\)
            \State \hspace{1cm} Reference element: $r$ 
            \State \hspace{1cm} Requested number of elements in the Sub-Pool: $size$ 
        \Function{CreateSubPool}{$U$, $r$, $size$}
            \While{\(k < size\)}
                \State mmr\_score $\gets$ 0
                \State sim\_element $\gets$ ""
                \For{$\alpha$ in $U$}
                    \State curr\_mmr\_score $\gets$ ComputeMMR($\alpha$, $r$, SubPool)
                    \If {curr\_mmr\_score > mmr\_score}
                        \State sim\_element $\gets$ $\alpha$
                        \State mmr\_score $\gets$ curr\_mmr\_score
                    \EndIf
                \EndFor
                \State \(k \gets k + 1\)
                \State SubPool.append(sim\_element)
            \EndWhile  
            \State \textbf{return} SubPool 
        \EndFunction
    \end{algorithmic}
\end{algorithm}

\begin{algorithm}
    \caption{BO-ICL algorithm}\label{alg:bo-icl}
    \begin{algorithmic}[1]
        \State \textbf{Initialize:}
        \State \hspace{1cm} Iteration counter: \(t \gets 0\)
        \State \hspace{1cm} Labeled dataset: \(\mathcal{L} \gets \{(x_n, y_n) | n \in {1,2,...,n_{\mathcal{L}}} \}\)
        \State \hspace{1cm} Unlabeled dataset: \(\mathcal{U} \gets \{(A_n) | n \in {1,2,...,n_{\mathcal{U}}}\}\)
        \State \hspace{1cm} Size of the SubPool: size
        \State \hspace{1cm} Size of the context: k
        \State \textbf{Input:} 
        \State \hspace{1cm} Surrogate function: \(f(x)\)\Comment{A LLM with long-term memory $\mathcal{M}$}
        \State \hspace{1cm} Acquisition function: \(\alpha(x)\)
        \Function{BO-ICL}{\(f(x)\), \(\alpha(x)\)}
            \State $\mathcal{M}_t$ $\gets$ $\mathcal{L}$
            \While{stopping criterion not met}
                \State \(t = t + 1\)
                \State \((x^+, y^+) = (\arg\max_{\mathcal{M}} \{y_n\}, max_{\mathcal{M}} \{y\})\)
                \State best $\gets (x^+, y^+)$ 
                \\
                \LComment{\(\textbf{Select next point:}\)}
                \State reference $\gets$ LLM($y^+$, k, $\mathcal{M}_{t-1}$)\Comment{Inverse design experiment generation. See Algorithm~\ref{alg:llm}}
                \State SubPool $\gets$ CreateSubPool($\mathcal{U}$, reference, size)\Comment{See Algorithm~\ref{alg:create-sub-pool}}
                \State \(r_{x,t} \gets \arg\max_{SubPool} \alpha(LLM(SubPool.x, k, \mathcal{M}_{t-1}))\)\Comment{Inference prediction generation. See Algorithm~\ref{alg:llm}}
                \\
                \LComment{\(\textbf{Evaluate objective:}\)}
                \State \( r_{y,t} \gets \text{Label obtained from the experiment}\)
                \If{\(r_{y,t} > \text{best}.r_y\)}
                    \State best \(\gets (r_x, r_y) = (r_{x,t}, r_{y,t})\)
                \EndIf
                \\
                \LComment{\(\textbf{Update LLM's memory:}\)}
                \State \(\mathcal{M}_t = \mathcal{M}_{t-1} \cup \{(r_{x,t}, r_{y,t})\}\)
            \EndWhile
            \State \textbf{Output:} best \Comment{Return best point point found}
        \EndFunction
    \end{algorithmic}
\end{algorithm}

\SIsection{Cost Analysis}\label{sec:cost-analysis}

To address monetary considerations associated with running BO-ICL, we present the cost and process parameters obtained directly from OpenAI for a complete run as presented in aforementioned results (Figure \ref{fig:bo-ocm}). This analysis considers the most expensive model used in this study (\texttt{gpt-4-0125-preview}) as well as the embedding model (\texttt{Ada-small}) [see Table \ref{tab:alloy_cost_table}]. The sub-sampling strategy illustrated in points A4–A8 of the flowchart [Figure \ref{fig:flowchart}] is central to achieving the observed low cost. The prediction and update steps incur costs only for output tokens, billed at \$10.00 per 1M output tokens, and involve 16 samples per iteration. Additionally, a single inverse design generation step contributes a cost corresponding to the number of tokens required to represent a single experimental procedure. Aside from this, there is a nominal one-time cost for embedding the initial pool ($\$0.1/1M$ token) after which the embedding representations are cached and reused throughout the design loop. This further supports using BO-ICL even with SOTA models at low cost deploy, contrary to reported mention \cite{Kristiadi2024-mf}.

\vspace{0.5cm} 
\begin{table}[h!]
\centering 
\begin{tabular}{c|c|c|c|c|c|c}
\textbf{Dataset} & \textbf{Samples} & \textbf{Replicates} & \textbf{Iterations} & \textbf{Inverse Design Model} & \textbf{Prediction Model} & \textbf{Total Cost} \\
\hline
AII & 9,000 & 5 & 30 & GPT-4o-preview & GPT-4o-preview & \$ 10.52 \\
OCM & 12,708 & 5 & 30 & GPT-4o-preview & GPT-4o-preview & \$ 12.22\\
RWGS & 3,720 & 2 & 6 & GPT-4o-preview & GPT-4o-preview & \$ 0.98\\

\end{tabular}
\caption{Cost evaluation for the Alloy, In-House RWGS, and OCM dataset}
\label{tab:alloy_cost_table}
\end{table}
\vspace{0.5cm}

\newpage

\SIsection{Prompts and system messages}\label{sec:sysmes}

The following system message led to better results and was used in every Bayesian optimization result that involved a chat model shown in this study:

\newpage

\begin{lstlisting}[breaklines=true, breakindent=0px, frame=single]
You are an expert in heterogeneous catalysis, with deep knowledge of the electronic properties of elements, especially how transition metals interact with various supports to synergistically catalyze reactions under different conditions. Your expertise extends to designing experimental procedures aimed at achieving desired material properties. Your capabilities are built on the most current and comprehensive information available, up to April 2023. When responding to inquiries, please focus on providing specific experimental procedures, without including explanations. Your approach should reflect a balanced integration of user-provided information and your base knowledge to identify the most impactful experimental strategies for their needs. Our goal is to deliver clear and direct procedural guidance, ENSURING that we match the user's example formats when responding, to identify the most effective experimental procedure for their needs. 
           
These are the possible parameters that can be input:
catalyst, name, precursor, support, M1b, M2b, M3b
Mn-Na2WO4/BN, Mn(NO3)2.6H2O, Na2WO4, BN, Mn(40), Na(40), W(20)
Mn-Na2WO4/MgO  Mn(NO3)2.6H2O, Na2WO4            MgO                               Mn (40)   Na (40)   W (20)
Mn-Na2WO4/Al2O3 Mn(NO3)2.6H2O, Na2WO4            Al2O3                             Mn (40)   Na (40)   W (20)
Mn-Na2WO4/SiC   Mn(NO3)2.6H2O, Na2WO4           SiC                               Mn (40)   Na (40)   W (20)
Mn-Na2WO4/SiCnf Mn(NO3)2.6H2O, Na2WO4           SiCnf                             Mn (40)   Na (40)   W (20)
Mn-Na2WO4/BEA    Mn(NO3)2.6H2O, Na2WO4          BEA                               Mn (40)   Na (40)   W (20)
Mn-Na2WO4/ZSM-5  Mn(NO3)2.6H2O, Na2WO4          ZSM-5                             Mn (40)   Na (40)   W (20)
Mn-Na2WO4/TiO2   Mn(NO3)2.6H2O, Na2WO4          TiO2                              Mn (40)   Na (40)   W (20)
Mn-Na2WO4/ZrO2   Mn(NO3)2.6H2O, Na2WO4          ZrO2                              Mn (40)   Na (40)   W (20)
Mn-Na2WO4/Nb2O5   Mn(NO3)2.6H2O, Na2WO4         Nb2O5                             Mn (40)   Na (40)   W (20)
Mn-Na2WO4/CeO2    Mn(NO3)2.6H2O, Na2WO4         CeO2                              Mn (40)   Na (40)   W (20)
Mn-Li2WO4/SiO2    Mn(NO3)2.6H2O, Li2WO4         SiO2                              Mn (40)   Li (40)   W (20)
Mn-MgWO4/SiO2     Mn(NO3)2.6H2O, MgWO4          SiO2                              Mn (50)   Mg (25)   W (25)
Mn-K2WO4/SiO2      Mn(NO3)2.6H2O, K2WO4         SiO2                              Mn (40)   K (40)    W (20)
Mn-CaWO4/SiO2      Mn(NO3)2.6H2O, CaWO4         SiO2                              Mn (50)   Ca (25)   W (25)
Mn-SrWO4/SiO2      Mn(NO3)2.6H2O, SrWO4         SiO2                              Mn (50)   Sr (25)   W (25)
Mn-BaWO4/SiO2      Mn(NO3)2.6H2O, BaWO4         SiO2                              Mn (50)   Ba (25)   W (25)
Mn-Li2MoO4/SiO2    Mn(NO3)2.6H2O, Li2MoO4       SiO2                              Mn (40)   Li (40)   Mo (20)
Mn-Na2MoO4/SiO2    Mn(NO3)2.6H2O, Na2MoO4       SiO2                              Mn (40)   Na (40)   Mo (20)
Mn-K2MoO4/SiO2     Mn(NO3)2.6H2O, K2MoO4        SiO2                              Mn (40)   K (40)    Mo (20)
Mn-FeMoO4/SiO2     Mn(NO3)2.6H2O, FeMoO4        SiO2                              Mn (50)   Fe (25)   Mo (25)
Mn-ZnMoO4/SiO2     Mn(NO3)2.6H2O, ZnMoO4        SiO2                              Mn (50)   Zn (25)   Mo (25)
Ti-Na2WO4/SiO2      Ti(OiPr)4, Na2WO4            SiO2                              Ti (40)   Na (40)   W (20)
V-Na2WO4/SiO2       VOSO4.xH2O (x = 3-5), Na2WO4 SiO2                              V (40)    Na (40)   W (20)
Fe-Na2WO4/SiO2      Fe(NO3)3.9H2O, Na2WO4        SiO2                              Fe (40)   Na (40)   W (20)
Co-Na2WO4/SiO2      Co(NO3)2.6H2O, Na2WO4        SiO2                              Co (40)   Na (40)   W (20)
Ni-Na2WO4/SiO2      Ni(NO3)2.6H2O, Na2WO4        SiO2                              Ni (40)   Na (40)   W (20)
Cu-Na2WO4/SiO2      Cu(NO3)2.5H2O, Na2WO4        SiO2                              Cu (40)   Na (40)   W (20)
Zn-Na2WO4/SiO2      Zn(NO3)2.6H2O, Na2WO4        SiO2                              Zn (40)   Na (40)   W (20)
Y-Na2WO4/SiO2       Y(NO3)3.6H2O, Na2WO4         SiO2                              Y (40)    Na (40)   W (20)
Zr-Na2WO4/SiO2      ZrO(NO3)2.2H2O, Na2WO4      SiO2                              Zr (40)   Na (40)   W (20)
Mo-Na2WO4/SiO2      (NH4)2MoO4, Na2WO4           SiO2                              Mo (40)   Na (40)   W (20)
Pd-Na2WO4/SiO2      Pd(OAc)2, Na2WO4             SiO2                              Pd (40)   Na (40)   W (20)
La-Na2WO4/SiO2      La(NO3)3, Na2WO4             SiO2                              La (40)   Na (40)   W (20)
Ce-Na2WO4/SiO2      Ce(NO3)3.6H2O, Na2WO4        SiO2                              Ce (40)   Na (40)   W (20)
Nd-Na2WO4/SiO2      Nd(NO3)3.6H2O, Na2WO4        SiO2                              Nd (40)   Na (40)   W (20)
Eu-Na2WO4/SiO2      Eu(NO3)3.5H2O, Na2WO4        SiO2                              Eu (40)   Na (40)   W (20)
Tb-Na2WO4/SiO2      Tb(NO3)3.5H2O, Na2WO4        SiO2                              Tb (40)   Na (40)   W (20)
Hf-Na2WO4/SiO2      Hf(OEt)4, Na2WO4             SiO2                              Hf (40)   Na (40)   W (20)
blank             -                                -                                 -         -         -         -
BN                -                                BN                                -         -         -         -
MgO               -                                MgO                               -         -         -         -
Al2O3             -                                Al2O3                             -         -         -         -
SiO2              -                                SiO2                              -         -         -         -
SiC               -                                SiC                               -         -         -         -
SiCnf             -                                SiCnf                             -         -         -         -
BEA               -                                BEA                               -         -         -         -
ZSM-5             -                                ZSM-5                             -         -         -         -
TiO2              -                                TiO2                              -         -         -         -
ZrO2              -                                ZrO2                              -         -         -         -
Nb2O5             -                                Nb2O5                             -         -         -         -
CeO2              -                                CeO2                              -         -         -         -
Na2WO4/SiO2       Na2WO4                           SiO2                              -         Na (67)  W (33)   -
Mn-WOx/SiO2       Mn(NO3)2.6H2O, (NH4)10H2(W2O7)6 SiO2                             Mn (67)   -         W (33)   -
Mn-MoOx/SiO2      Mn(NO3)2.6H2O, (NH4)2MoO4      SiO2                              Mn (67)   -         Mo (33)  -
Mn-Na/SiO2        Mn(NO3)2.6H2O, NaNO3           SiO2                              Mn (50)   Na (50)   -
WOx/SiO2          (NH4)10H2(W2O7)6                SiO2                              -         -         -         W (100)
Na/SiO2           NaNO3                            SiO2                              -         Na (100) -         -

Your experimental procedures can only use those parameters. This is more information to help control how you output these procedures: The metal loadings to a unit gram of support were fixed at 0.371 mmol for Metal 1, 0.370 or 0.185 mmol for Metal 2 (depending on the valence), and 0.185 mmol for Metal 3. The values in parentheses need to correspond to relative atomic percentages of M1-M3: to a unit gram of the support, only choose from, 0.371 mmol M1, 0.370 or 0.185 mmol M2, and 0.185 mmol M3 or 0.0. Use the exact format of the given examples when responding, given a property like C2 yield.
\end{lstlisting}

\SIsection{Datasets}\label{sec:datasets}

To validate this workflow, we focus on application areas of global significance, such as catalytic material design for green house gas (GHG) upcycling. Accelerating material discovery in this area can reduce reliance on crude oil used for high-demand chemical precursors by using relevant waste C1 species like $CO_2$, to close the emission loop and life cycle\cite{Duyar2015-gs}. Catalytic materials can play a major role in promoting each step in such a circular carbon economy by helping to offset related cost. These materials allow one to selectively exploiting the hysteresis gap in relevant thermodynamically controlled chemical processes. For example, inorganic catalytic materials are well studied to selectively reduce $CO_2$ (C=O bond separation energy of 432 kj/mol) for rapid conversion of CO to valuable $C_2$ to $C_8$ products ($CO_2$ Fischer-Tropsch synthesis). The size of the design space for these materials, the necessity to match relevant reaction conditions for efficacy, and the cost associated with running relevant experiments makes it an ideal test space to test capabilities of using frozen SOTA LLMs as useful surrogate models. 

Extrapolating surface temperature fluctuations in response to the average $CO_2$ concentration variation over the last decade, at current carbon dioxide removal (CDR) efficiencies of 1.3 million tonnes of $CO_2$ per year, annual capture capacities must increase by a factor of 30, by 2030, to meet UN targets and avoid predicted catastrophic affects of global warming\cite{Dunne2024-rn}. Thus, without significant economic incentive, the widespread adoption rate of necessary capture and conversion processes may not match the pace of greenhouse gas accumulation in the troposphere. Traditional catalytic discovery and deployment pathways have a record for broad implementation time-lines ranging from 5-40 years. This rate is unacceptable if the goal is to find materials that will allow us to avoid adverse affects due to tropospheric temperature fluctuations. 

Although catalyst informatics is a rapidly growing field, due to its potential to increase efficiency in the material discovery process, a common challenge in developing models useful for structure-property approximations is the disproportionate availability of experimental data, relative to the size of the parameter design space. Using language as an agnostic feature space with BO may offer a solution to this issue as it could allow the unbiased use of material data for design and property prediction, while significantly reducing the cost of experiments needed to identify effective materials. In this setting, we can use BO-ICL to efficiently guide us through the complex design space of matching material and process design, by directly representing materials as standard operating procedures that includes important levers for both synthesizing and testing these materials.

\subsection{Solubility}\label{ssec:solubility}

The Estimated Solubility (ESOL)\cite{Delaney2004-de} dataset is a widely used benchmark in cheminformatics for predicting the aqueous solubility of small organic molecules. 
It consists of a collection of experimentally measured solubility values expressed in log molar units (logS).
Originally, ESOL is published with the SMILES\cite{Weininger1988-gy} representation of the molecules and the LogS values.
This study used the PubChem API\cite{Kim2018-zz} to get IUPAC names.
IUPAC names were input to our LLM models, and embedded representations of such names were used for the baselines.

\subsection{Oxidative Coupling of Methane}\label{ssec:ocm}

\begin{figure}[H]
    \centering
    \includegraphics[width=0.6\linewidth]{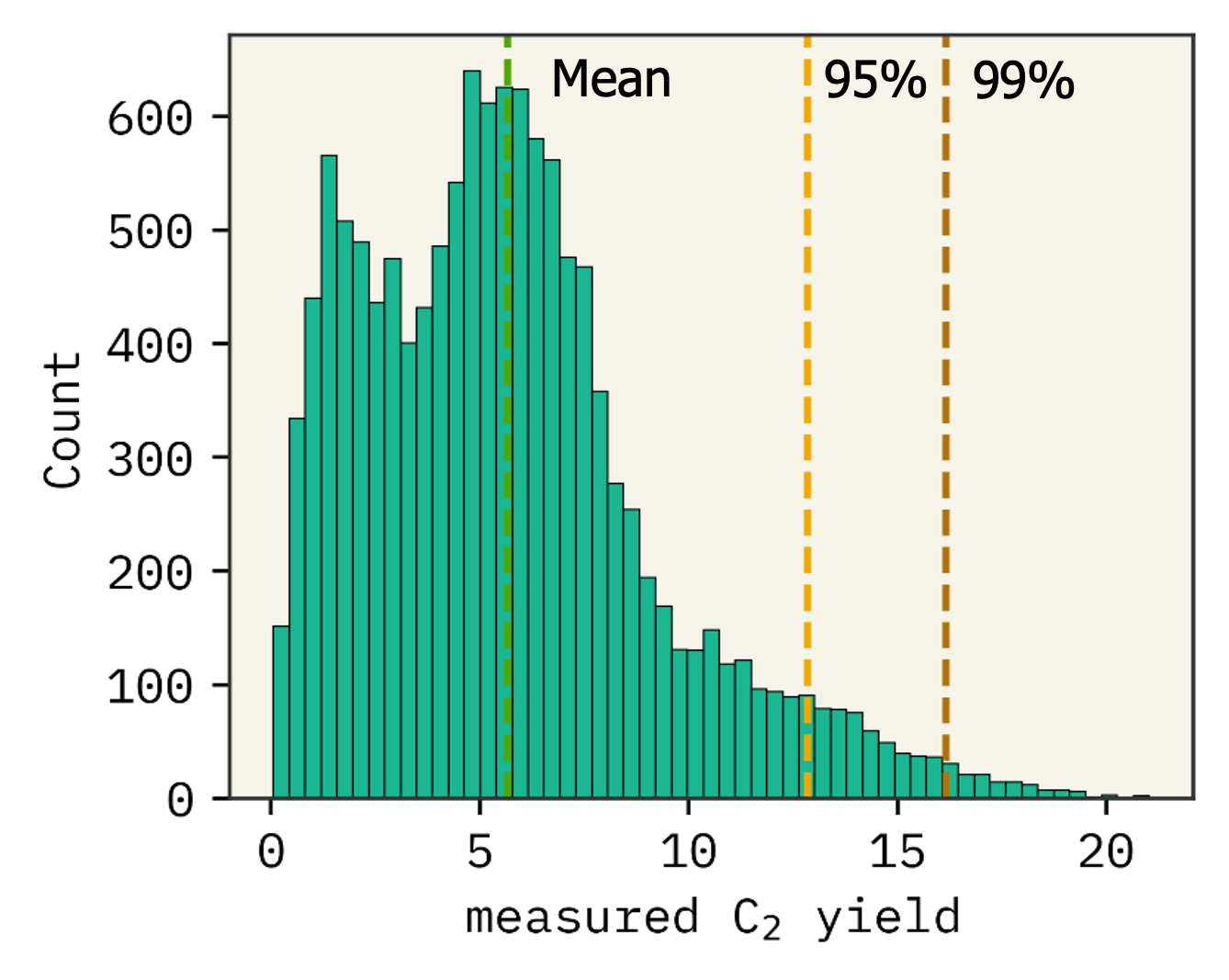}
    \caption{Histogram of OCM dataset performance distribution for $C_2$ yield (\%) with annotated quantiles: Mean (green), $95^{th}$ percentile (orange), $99^{th}$ percentile (brown)}
    \label{fig:ocm_dist}
\end{figure}

\begin{figure}[H]
    \centering
    \includegraphics[width=1.1\linewidth]{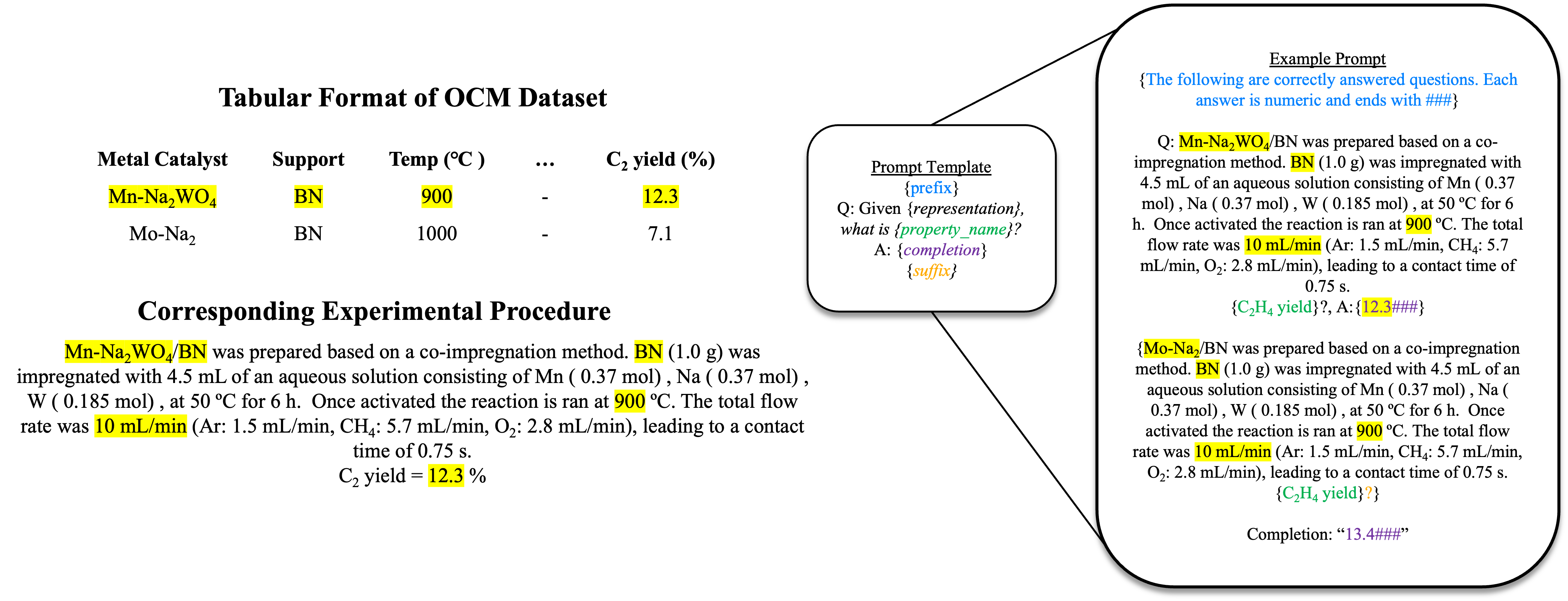}
    \caption{Example of original OCM tabular data and corresponding natural language representation used in BO-ICL evaluations. Additionally, a depiction of a prompt with $k=1$, offering a single context example at inference.}
    \label{fig:tab_text}
\end{figure}

This dataset focuses on catalysis optimization for the oxidative coupling of methane (Equation \ref{oxi-meth}). \citet{Nguyen2020-kf} evaluated 12,708 experimental configurations across a range of parameters, including different catalyst active phases, support types, chemical compositions, reaction temperature, and reactant contact times. \citet{Nguyen2020-kf} reported the catalyst performance for $C_2$ (\%) yield production under oxidative coupling of methane reaction conditions for 59 different catalysts, including reference materials. Catalyst performance was measured using a high-throughput screening instrument for consistent analyses, resulting in a high-fidelity dataset, ideal for early testing of BO-ICL. Reported tabulated conditions and results were converted to natural language representations (e.g. Figure \ref{fig:tab_text})\cite{Dinh2022-oj}. The property value distribution is visible in Figure \ref{fig:ocm_dist} as a histogram, highlighting the sparsity in performance configurations above a 15\% yield.

\begin{equation}
\text{(OCM)} \quad 2\mathrm{CH}_4 + \mathrm{O}_2 \rightarrow \mathrm{C}_2\mathrm{H}_4 + 2\mathrm{H}_2\mathrm{O} \quad \Delta H_{rxn} = -280 \, \text{kJ/mol}
\label{oxi-meth}
\end{equation}

\subsection{Alloy Interface}\label{ssec:alloy}

\begin{figure}[H]
    \centering
    \includegraphics[width=0.8\linewidth]{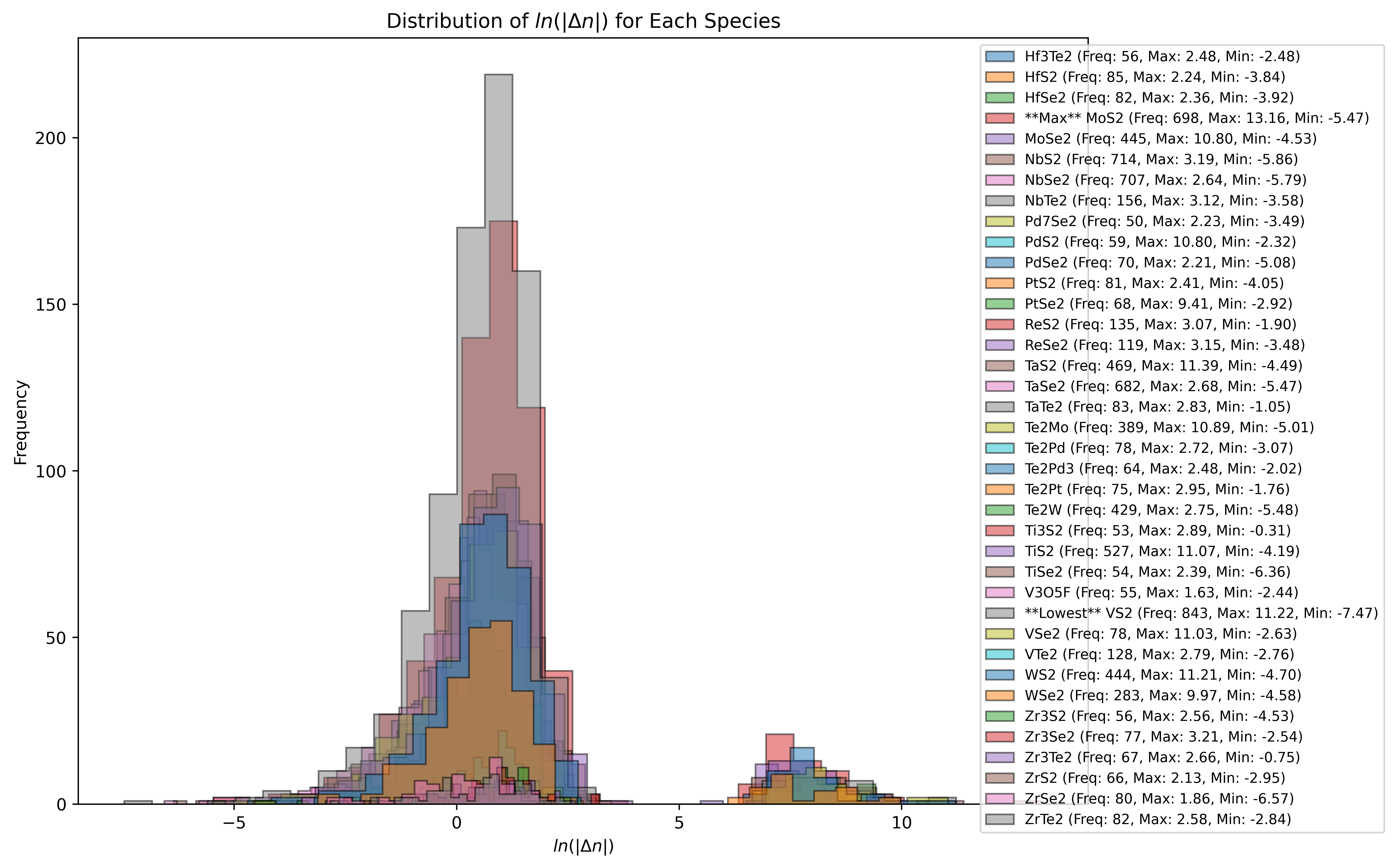}
    \caption{AII dataset distribution with color coded alloys and a range of their associated charge transfer values.}
    \label{fig:alloy_dist}
\end{figure}

This dataset was forged using data from \cite{Gerber2023-xl}, to calculate the charge transfer between two alloy interfaces modeled as a parallel plate capacitor. In this paper, an equilibrium Fermi-level $E{'}_{f}$ was used to analytically solve for the charge transfer vector. Here, we leave it up to the BO-ICL to realize this relationship only given the Fermi level of the two alloys obtained separately from the materials project, the interface model (e.g. capacitor), and the transfer distance $d$, set as the sum of the largest van der Waals radii in each alloy. The log scaled distribution of charge transfer values and alloy combinations are visible in SI Figure \ref{fig:alloy_dist}.
\begin{equation}
\int_{E'_{f1}}^{E_{f1}} dE \, g_1(E) = \int_{E'_{f2}}^{E_{f2}} dE \, g_2(E)
\end{equation}

\begin{equation}
e \Delta n = \epsilon_0 \frac{E'_{F}}{d}
\end{equation}

\subsection{In-house RWGS}\label{ssec:rwgs}

\begin{figure}[H]
    \centering
    \includegraphics[width=0.4\linewidth]{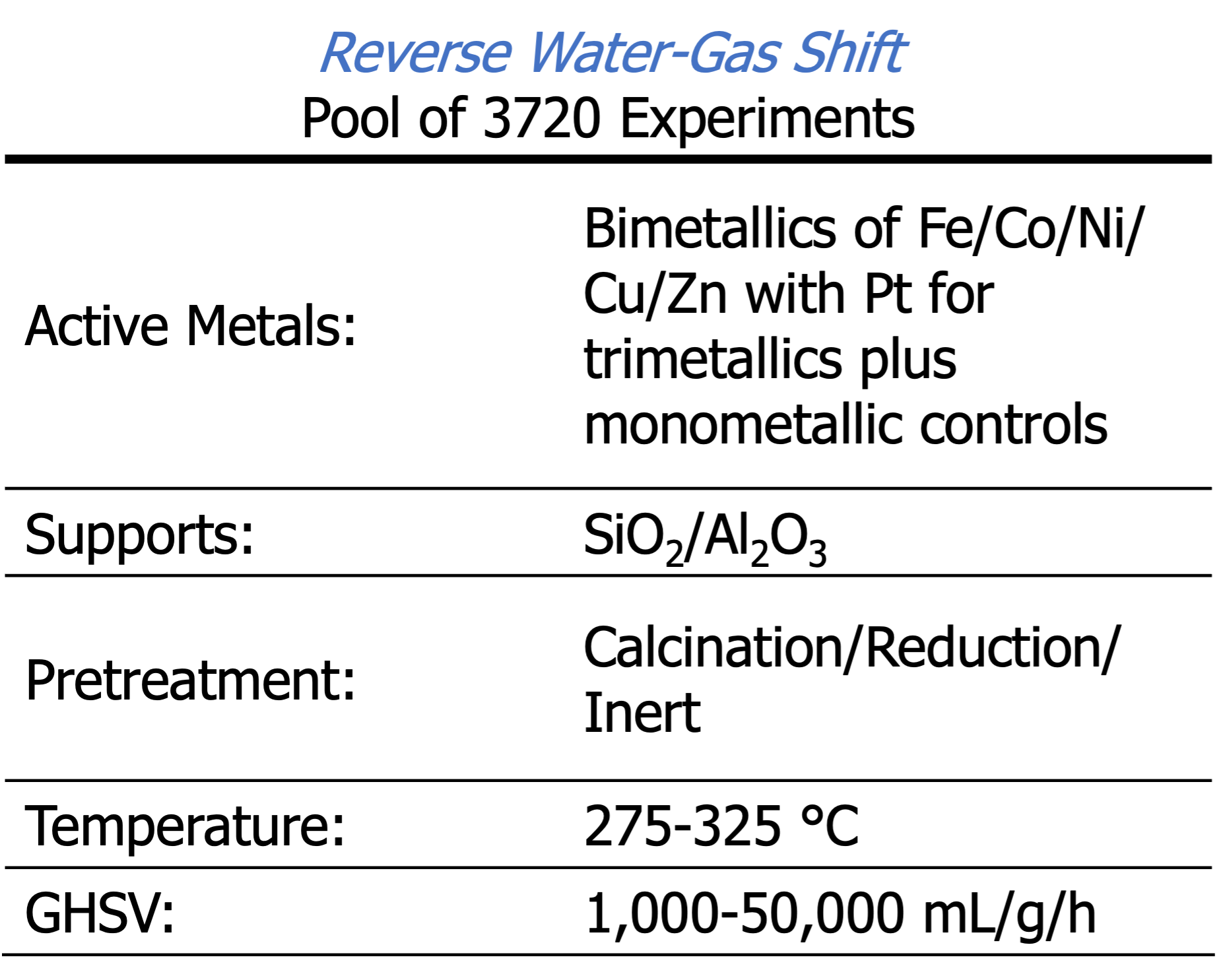}
    \caption{Reverse water-gas shift experimental pool. All Reactions were run at a total pressure of 1 atm}
    \label{fig:rwgs}
\end{figure}

To further verify BO-ICL's application potential, we synthesized a custom pool of 3720 possible experiments aimed at identifying catalytic materials favorable for selective CO production under reverse water-gas shift (RWGS) reaction conditions. This experimental setup, which involves iterative human-led synthesis and reactor configuration, allows us to assess BO-ICL’s performance when working with data that contains aleatoric uncertainty, typical of real-world catalysis studies. We constrained the material design space by limiting earth-abundant transition metal impregnation to 4.9 wt. \% and .1 wt. \% of platinum on supports that offer reaction-appropriate thermal stabilities and high surface area ($SiO_{2}/\gamma-Al_{2}O_{3}$), visible in Figure: \ref{fig:rwgs}. This sample space offers interesting considerations for such an application for reasons beyond $CO_2$ remediation solutions. For example, the choice of exploring equilibrium limited endothermic reaction offers an upper limit on $CO$ yield, governed by thermodynamics, and thus a clear termination policy is available. Further, the depth of available knowledge surrounding this reaction also allows us to introduce controlled complexities into the catalytic system. We know that incorporating low concentrations of precious metals like Pt can rapidly dissociate $H_2$ due to the insignificant activation barrier for $H_2$ adsorption on Pt metal surfaces \cite{Yu2012-ze}. By introducing Pt, we can further shift the focus toward optimizing selective $CO_2$ activation chemistries, by eliminating concerns of active hydrogen availability. This shift amplifies the need for finer process parameter tuning to match favorable synergistic binding energies that will resist deep reduction into thermodynamically favorable, and undesired, methane formation.

\subsubsection{Thermodynamic upper limit approximation for $CO_{yield}$}:

\begin{equation}
\text{(RWGS)} \quad \text{CO}_2 + \text{H}_2 \rightleftharpoons \text{CO} + \text{H}_2\text{O} \quad 
\end{equation}

\begin{center}
\begin{tabular}{c|c|c|c|c}
 & \(\text{CO}_2\) & \(3\text{H}_2\) & \(\text{CO}\) & \(\text{H}_2\text{O}\) \\
\hline
\textbf{Initial} & \(n_0\) & \(3n_0\) & - & - \\
\textbf{Change} & \(-n_0x\) & \(-n_0x\) & \(+n_0x\) & \(+n_0x\) \\
\textbf{Equilibrium} & \(-n_0(1-x)\) & \(-n_0(3-x)\) & \(n_0x\) & \(n_0x\) \\
\end{tabular}
\end{center}

Here, \( n_0 \) represents the initial moles of \(\text{CO}_2\), \( x \) denotes the conversion of \(\text{CO}_2\), and \( K_{\text{eq}} \) is the equilibrium constant.

\begin{equation}
K_{\text{eq}} = \frac{[\text{CO}][\text{H}_2\text{O}]}{[\text{CO}_2][\text{H}_2]} = \frac{x^2}{(1-x)(3-x)} = \exp\left(-\frac{\Delta G^\circ}{RT}\right)
\end{equation}

Where:
$\Delta G^\circ$ represents the standard Gibbs free energy change (\(\text{J/mol}\)),
$R$ is the universal gas constant (\(8.314 \, \text{J/mol·K}\)), and 
$T$ means the Temperature in Kelvin (K).

At equilibrium, at \( T = 325 \, ^\circ\text{C} \) and \( P = 1 \, \text{atm} \), the conversion of \(\text{CO}_2\), denoted as \( x_{\text{CO}_2} \), is approximately 26\%.

The yield of \(\text{CO}\) can be calculated as:
\begin{equation}
\text{CO}_{\text{yield}} = x_{\text{CO}_2} \cdot S_{\text{CO}_2}
\end{equation}

Where:
$x_{\text{CO}_2}$ is the conversion of $\text{CO}_2$, defined as the fraction of $\text{CO}_2$ reacted, calculated as:
\begin{equation}
    x_{\text{CO}_2} = \frac{\text{CO}_2^{\text{in}} - \text{CO}_2^{\text{out}}}{\text{CO}_2^{\text{in}}} \cdot 100\%
\end{equation}
    
    $S_{\text{CO}_2}$ means the selectivity for $\text{CO}$ formation, defined as the ratio of the moles of $\text{CO}$ formed to the moles of $\text{CO}_2$consumed:
\begin{equation}
    S_{\text{CO}_2} = \frac{\text{CO}^{\text{out}}}{\text{CO}_2^{\text{in}} - \text{CO}_2^{\text{out}}}\cdot 100\%.
\end{equation}

\subsubsection{Catalyst Synthesis and Testing}

For synthesizing materials for the RWGS dateset, respective nitrate precursors were first  proportionally and separately dissolved in Mill Q water for Fe, Co, Ni, Cu, Zn, and Pt metals. Then the appropriate concentrations of dissolved metal precursors were mix into a single beaker to ensure a total loading of 5 \% wt. of metal loading with respect to both the weight of the support and transition metals. We then add alloy solution drop-wise to the selected support using the incipient wetness impregnation synthesis method. The impregnated catalyst was then dried, by ramping the system temperature to 90 \textdegree C at 1 \textdegree C per minute with hold time of 4 hours for gentle water removal. Following this, another temperature ramp at the same rate to 450 \textdegree C and calcined, when called for, in air for 4 hours.

\subsubsection{Reactor Tests}

Each catalyst was loaded into a stainless-steel reactor (outer diameter: 6.35 mm; inner diameter: 4.57 mm; reactor length: 40 cm). When reduced, 40 mL/min of H\textsubscript{2} was introduced for 2 hours at 450 \textdegree C and 50 psig, or alternatively, the catalyst was degassed under 20 mL/min of Ar for 2 hours. Following reduction or degassing, the reactor was isolated, and bypass effluents, under 14.7 psig of pressure, were analyzed to establish a baseline. The gas composition for the reverse water-gas shift (RWGS) reaction consisted of 10 mL/min CO\textsubscript{2}, 30 mL/min H\textsubscript{2}, and 5 mL/min Ar, resulting in a H\textsubscript{2}/CO\textsubscript{2} ratio of 3:1. Catalyst mass loadings were varied to achieve gas hourly space velocities (GHSV) ranging from 1,000 to 50,000 mL/g/h.

The Gas Hourly Space Velocity (GHSV) is calculated as:

\begin{equation}
\text{GHSV} = \frac{F}{V_c}
\end{equation}

where \( F \) is the volumetric flow rate of gas (e.g., in mL/h), and \( V_c \) is the estimated catalyst bed volume. Isothermal reactions were run for 8 h. Effluent reactor concentrations were analyzed by an in-line Agilent Technologies 7890B GC system
equipped with a flame ionization detector (FID) and a thermal
conductivity detector (TCD). The concentration of each gas phase species was calibrated by correlating the peak area of the pure compound to its concentration in a calibration gas
standard. For all reactor experiments, the carbon balance closes to 100\% ± 2\%.

\subsubsection{RWGS optimization results}

\begin{table}[h!]
\centering
\small 
\setlength{\tabcolsep}{4pt} 
\caption{In-House Dataset Results}
\label{tab:inhouse-results}
\begin{tabular}{c|ccc|ccc|ccc}
    \hline
    \multirow{3}{*}{\shortstack{\textbf{Iteration}}} 
    & \multicolumn{3}{c|}{\cellcolor{green!20}\textbf{\texttt{gpt-4-32k-0314}}} 
    & \multicolumn{3}{c|}{\cellcolor{orange!40}\textbf{\texttt{gpt-4-0125-preview}}} 
    & \multicolumn{3}{c}{\cellcolor{blue!60!red!35!white!60}\textbf{\texttt{random walk}}} \\
    & \multicolumn{3}{c|}{\small(Greedy)} 
    & \multicolumn{3}{c|}{\small(UCB)} 
    & \multicolumn{3}{c}{\small} \\
    & \textbf{Catalyst} & \textbf{Yield$_{CO}$} & \textbf{Temp}  
    & \textbf{Catalyst} & \textbf{Yield$_{CO}$} & \textbf{Temp}  
    & \textbf{Catalyst} & \textbf{Yield$_{CO}$} & \textbf{Temp}  \\
    &  & (\%) & (\textdegree C) 
    &  & (\%) & (\textdegree C) 
    &  & (\%) & (\textdegree C) \\
    \hline
    i& $^n$Co-Zn-Pt/SiO$_2$ & 1.0 & 325 & $^n$Co-Zn-Pt/SiO$_2$ & 1.0 & 325 & $^n$Co-Zn-Pt/SiO$_2$ & 1.0 & 325 \\
    1& $^n$Co-Zn-Pt/SiO$_2$ & 1.0 & 275 & Fe/SiO$_2$ & 3.7 & 275 & *Fe-Zn/SiO$_2$ & 1.3 & 300 \\
    2& Co-Zn-Pt/SiO$_2$ & 1.5 & 300 & $^n$Ni-Cu/Al$_2$O$_3$ & 5.3 & 325 & Fe-Zn/SiO$_2$ & 1.5 & 300 \\
    3& Co-Cu-Pt/SiO$_2$ & 5.4 & 300 & $^n$Co-Zn-Pt/Al$_2$O$_3$ & 22.0 & 325 & Fe-Zn/Al$_2$O$_3$ & 4.3 & 300 \\
    4& Co-Cu-Pt/SiO$_2$ & 8.2 & 325 & $^*$Co-Pt/Al$_2$O$_3$ & 18.5 & 325 & Cu/SiO$_2$ & 2.5 & 325 \\
    5& Co-Pt/SiO$_2$ & 25.9 & 325 & $^*$Co-Zn-Pt/SiO$_2$ & 17.0 & 325 & $^{n*}$Co-Ni-Pt/Al$_2$O$_3$ & 6.2 & 300 \\
    \hline
\end{tabular}
\caption*{\textbf{Reaction conditions}: $P = 0.101$ MPa, GHSV = 36 L h$^{-1}$ g$^{-1}$, H$_2$:CO$_2$ ratio = 3, Superscripts: * - Ar Pre-treatment, $n$ - non-calcined, Note: If a sample was not Pre-treated in Ar, it was reduced using $H_2$ at 450 \textdegree C at 50 psi, prior to flowing RWGS reactants.}
\end{table}

\newpage
\SIsection{Baselines}\label{sec:baselines}

\subsection{Analytical random}\label{ssec:an_rand}

All Bayesian optimization plots show $y_N^*$ -- the current best at sample count $N$. 
The random baseline was estimated via a quantiling of the data points. 
Namely, for random sampling $y_N^*$ is estimated with:

\begin{equation}
\begin{split}
    E[y_N^*] &= \sum_i^K y_i\textrm{P}(s_m = y_i)\;;\; s_m = \textrm{max}\left(y_1, y_2,\ldots, y_N\right) \\
    &\approx \sum_j^Q\left(j^N - (j - 1)^N\right)\left(\frac{1}{Q}\right)^N q_j
\end{split}
\end{equation}

where $q_i$ is the $i$th quantile of $y$ (out of $Q$) and $K$ is the number of datapoints in the pool. 

\begin{figure}[h!]
    \centering
    \includegraphics[width=0.5\linewidth]{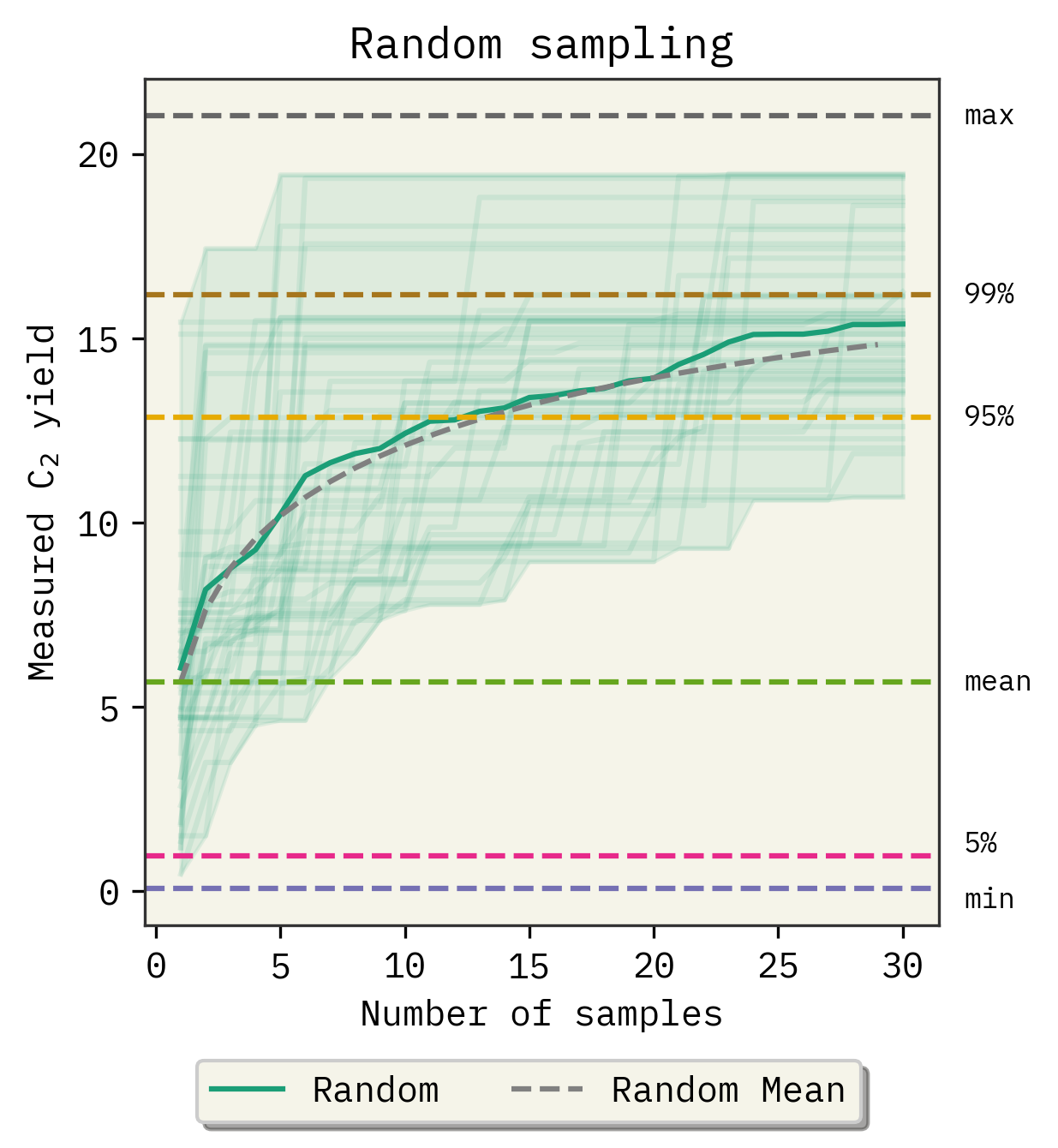}
    \caption{Comparison between a large number of random sampled trajectories with the analytical derived expected value of a random sampling. The analytical values are shown as a gray dashed line while the random sample is shown in green. Each independent trajectory is shown as shaded green lines and the average is displayed as the bold green line. Within 50 independent runs, the average of the random sampling converges to the analytical curve.}
    \label{fig:bo-random}
\end{figure}

To ensure our modeling was correct, we proof case this equation by comparing it with the average of several independent random sampled trajectories. Figure~\ref{fig:bo-random} shows a plot with both the analytical and the average of 50 random runs. As expected, the expected values for each number of samples are almost identical.

\subsection{OCM dataset with no true correlation}\label{ssec:rand_label}

\begin{figure}[hbt!]
    \centering
    \includegraphics[width=0.5\linewidth]{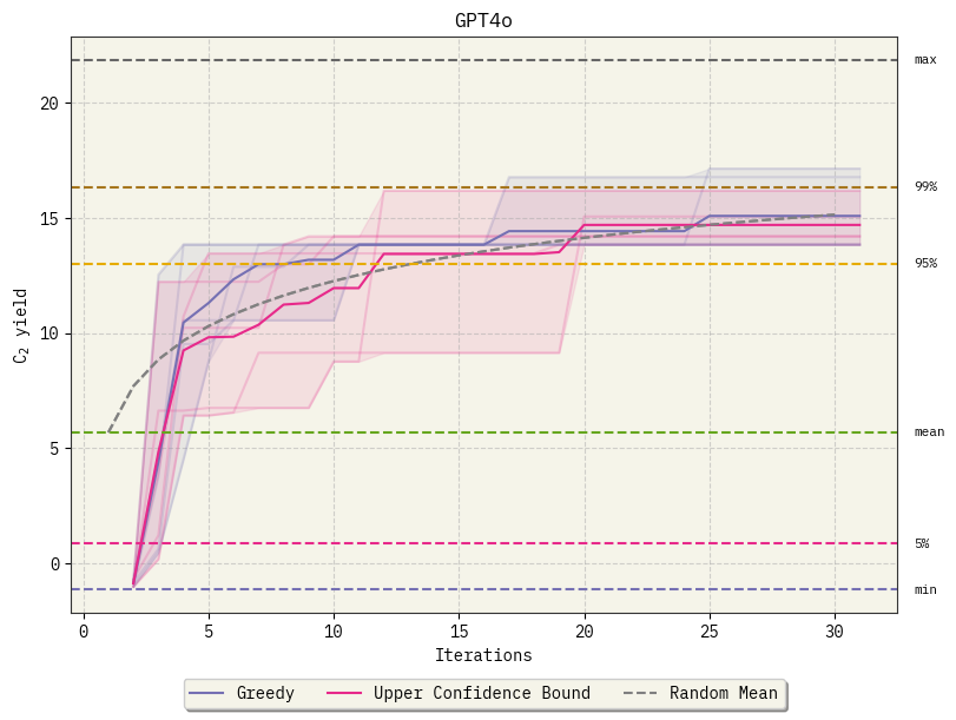}
    \caption{BO-ICL trajectory on corrupted OCM. This experiment shows that without a true correlation between the input experimental procedure, BO-ICL will be ineffective in guiding optimization within the design space.}
    \label{fig:ocm_false_labels}
\end{figure}

\begin{figure}[hbt!]
    \centering
    \includegraphics[width=0.5\linewidth]{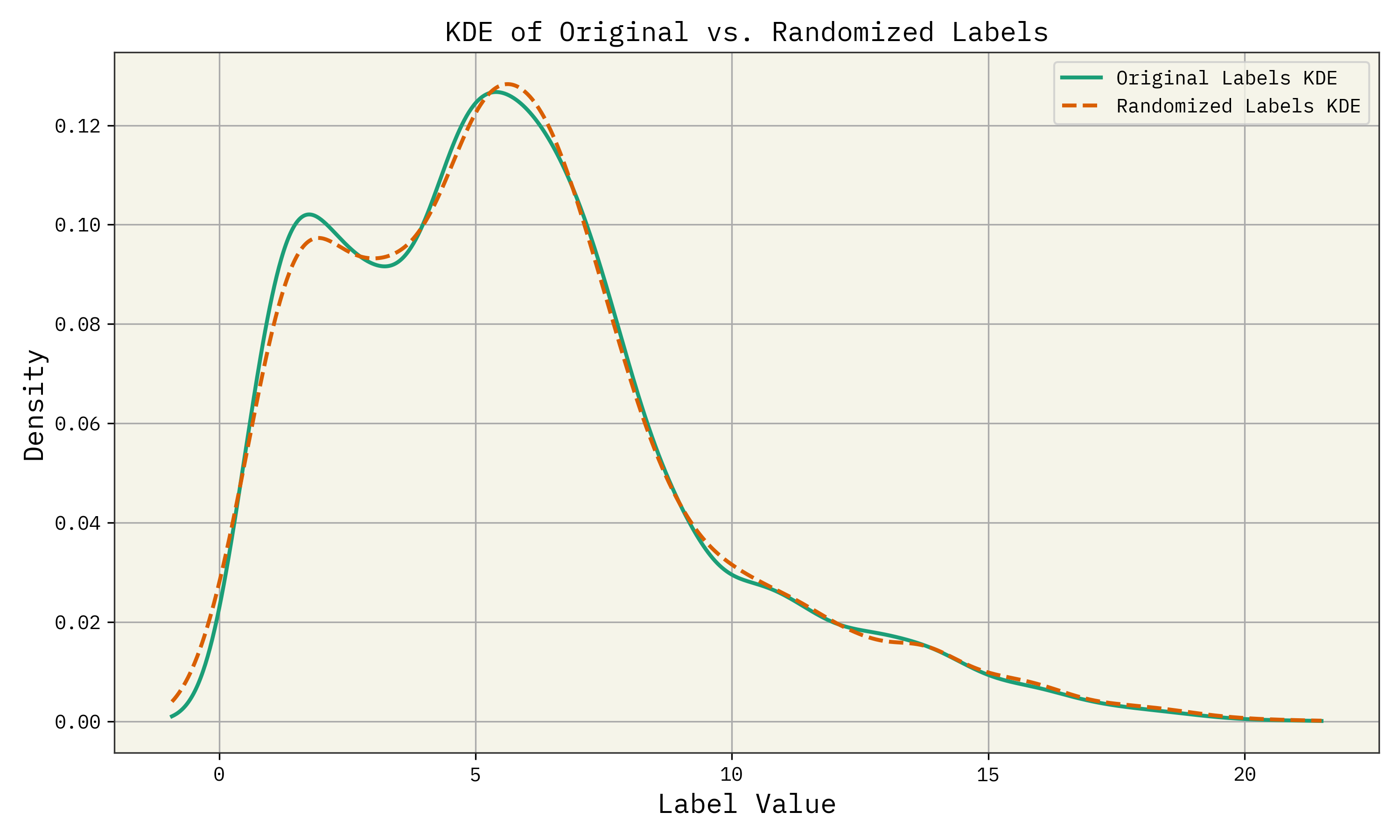}
    \caption{Comparison of probability distributions between the original and randomized OCM labels using Gaussian KDE. The randomized labels were generated by sampling from the KDE-fitted distribution of the original labels and then shuffling the samples. This preserves the overall label distribution while removing any correlation with the input features.}
    \label{fig:ocm_false_label_distribution}
\end{figure}

To demonstrate that the correlation between the chemical information in the input paragraphs and the target values is essential for optimization using BO-ICL, we artificially corrupted the OCM dataset. Specifically, we used kernel density estimation (via gaussian\_kde) to approximate the probability distribution of the original labels, as shown in. From this estimated distribution, we sampled a new set of labels with the same cardinality as the original dataset. These sampled labels were then randomly shuffled to preserve the overall distribution while eliminating any potential correlation between the input procedural features and the labels. This process yields a randomized label set that retains the original distribution's shape, visible in Figure~\ref{fig:ocm_false_label_distribution}, but has no functional relationship to the inputs. Figure~\ref{fig:ocm_false_labels} shows that upon corrupting the dataset, BO-ICL with \texttt{gpt-4o} can only perform equally to random sampling the dataset.

\subsection{k-Nearest neighbor}\label{ssec:knn}
The k-Nearest Neighbors (knn) model uses a space of embedded strings to perform the similarity search. It stores embeddings using the OpenAI embed model \texttt{text-ada-embedding-002} computed from the string with the experimental procedure description. 
For inference, it retrieves $k$ experiments with the greatest cosine similarity score from this saved space. The prediction is then the average of the $k$ retrieved experiments.

\subsection{Kernel ridge regression}\label{ssec:krr}
The kernel ridge regression (krr)\cite{Saunders1998-dp, Vu2015-nt} was implemented from scratch using numpy. The goal of krr is to minimize the following loss function:

\begin{equation}
    L((\boldsymbol{x}, y) | w)= \sum_{i=1}^{n}\left( y_i - w^{T}x_i \right)^2 + \lambda||w||_2
\end{equation}

In our implementation, we compute the kernel trick as

\begin{equation}
    \boldsymbol{K} = k(x_i, x_j) = \phi(x_i)^{T}*\phi(x_j)
\end{equation}

where, the function $\phi$ is the embedding function.
In our application, the embeddings are normalized after calculation.
The training involves solving the linear equation for $\alpha$, where $\alpha$ is the tensor of coefficients for the model.

\begin{equation}
    y = \boldsymbol{K}\alpha + \lambda\alpha
\end{equation}

On inference time, the following equation is computed.

\begin{equation}
    \hat{y} = \sum_{i=1}^{n}\alpha\phi(x_i)^{T}\phi(x'_j)
\end{equation}

\subsection{Gaussian process}\label{ssec:gpr}

Gaussian Process (GP) is a probabilistic model often used to model the unknown objective function \( f \) on a Bayesian optimization. The GP is a non-parametric model that defines a prior distribution over functions. Detailed discussion of GPs can be found in \citet{Rasmussen2005-xi} and \citet{Frazier2018-hq}.
Briefly, given a set of observations \( \mathcal{D} = \{(x_i, y_i)\}_{i=1}^n \), where \( y_i = f(x_i) + \epsilon \) and \( \epsilon \sim \mathcal{N}(0, \sigma^2) \) represents noise in the function evaluations, the GP models the objective function as:

\begin{equation}
f(x) \sim \mathcal{GP}(\mu(x), k(x, x'))
\end{equation}

where \( \mu(x) \) is the mean function and \( k(x, x') \) is the covariance function or kernel, encoding the similarity between points \( x \) and \( x' \).

The posterior distribution of the function at any new point \( x_* \), after observing data \( \mathcal{D} \), is given by:

\begin{equation}
\mu_*(x_*) = \mu(x_*) + k(x_*, X) K^{-1} (y - \mu(X))
\end{equation}
\begin{equation}
\sigma^2_*(x_*) = k(x_*, x_*) - k(x_*, X) K^{-1} k(X, x_*)
\end{equation}

where \( X \) is the set of observed points, \( K \) is the covariance matrix with elements \( K_{ij} = k(x_i, x_j) \), and \( k(x_*, X) \) is the vector of covariances between the new point \( x_* \) and the observed points \( X \).

In this study, GP was implemented using the bo-torch\cite{balandat2020botorch} package. More specifically, we used the \texttt{SingleTaskGP} as the regressor, which uses the Matérn-5/2 kernel. To prepare the data, we computed embeddings using the OpenAI embed model \texttt{text-ada-embedding-002}. Ada embeddings is a vector of 1532 dimensions. To simplify the GP training, we used an isomap to reduce the input dimension from 1532 to 32.

\SIsection{Additional results}\label{sec:reg_results}

\begin{table}[h!]
\centering
\caption{Statistical tests}\label{tab:pvalue}
\begin{tabular}{clcc}
    Dataset & Model & T-test & p-value \\\hline
    \multirow{16}{*}{ESOL}& gpt-3.5-turbo-0125 & k=1 $\leftrightarrow$ k=2 & 0.01719\\
                          & gpt-3.5-turbo-0125 & k=2 $\leftrightarrow$ k=5 & 0.19953 \\
                          & gpt-3.5-turbo-0125 & k=5 $\leftrightarrow$ k=10 & 0.02413\\\cline{2-4}
                          & gpt-3.5-turbo-0125 & T=0.05 $\leftrightarrow$ T=0.1 & 0.04135\\
                          & gpt-3.5-turbo-0125 & T=0.1 $\leftrightarrow$ T=0.5 & 0.15216\\
                          & gpt-3.5-turbo-0125 & T=0.5 $\leftrightarrow$ T=0.7 & 0.00596\\
                          & gpt-3.5-turbo-0125 & T=0.7 $\leftrightarrow$ T=1.0 & 0.58761\\
                          & gpt-3.5-turbo-0125 & T=1.0 $\leftrightarrow$ T=1.5 & 0.04197\\\cline{2-4}
                          & gpt-3.5-turbo-0125 & N=1 $\leftrightarrow$ N=5 & 0.29242\\
                          & gpt-3.5-turbo-0125 & N=5 $\leftrightarrow$ N=10 & 0.06123\\
                          & gpt-3.5-turbo-0125 & N=10 $\leftrightarrow$ N=25 & 0.41665\\
                          & gpt-3.5-turbo-0125 & N=25 $\leftrightarrow$ N=50 & 0.00475\\
                          & gpt-3.5-turbo-0125 & N=50 $\leftrightarrow$ N=100 & 0.35680\\
                          & gpt-3.5-turbo-0125 & N=100 $\leftrightarrow$ N=250 & 0.02949\\
                          & gpt-3.5-turbo-0125 & N=250 $\leftrightarrow$ N=500 & 0.03278\\
                          & gpt-3.5-turbo-0125 & N=500 $\leftrightarrow$ N=1k & 0.03216\\\hline 
    \multirow{16}{*}{OCM} & gpt-3.5-turbo-0125 & k=1 $\leftrightarrow$ k=2 & 0.01196\\
                          & gpt-3.5-turbo-0125 & k=2 $\leftrightarrow$ k=5 & 0.19324 \\
                          & gpt-3.5-turbo-0125 & k=5 $\leftrightarrow$ k=10 & 0.98559\\\cline{2-4}
                          & gpt-3.5-turbo-0125 & T=0.05 $\leftrightarrow$ T=0.1 & 0.01115\\
                          & gpt-3.5-turbo-0125 & T=0.1 $\leftrightarrow$ T=0.5 & 0.13083\\
                          & gpt-3.5-turbo-0125 & T=0.5 $\leftrightarrow$ T=0.7 & 0.10318\\
                          & gpt-3.5-turbo-0125 & T=0.7 $\leftrightarrow$ T=1.0 & 0.61822\\
                          & gpt-3.5-turbo-0125 & T=1.0 $\leftrightarrow$ T=1.5 & 0.01191\\\cline{2-4}
                          & gpt-3.5-turbo-0125 & N=1 $\leftrightarrow$ N=5 & 0.18649\\
                          & gpt-3.5-turbo-0125 & N=5 $\leftrightarrow$ N=10 & 0.02414\\
                          & gpt-3.5-turbo-0125 & N=10 $\leftrightarrow$ N=25 & 0.02752\\
                          & gpt-3.5-turbo-0125 & N=25 $\leftrightarrow$ N=50 & 0.95171\\
                          & gpt-3.5-turbo-0125 & N=50 $\leftrightarrow$ N=100 & 0.16021\\
                          & gpt-3.5-turbo-0125 & N=100 $\leftrightarrow$ N=250 & 0.31569\\
                          & gpt-3.5-turbo-0125 & N=250 $\leftrightarrow$ N=500 & 0.14796\\
                          & gpt-3.5-turbo-0125 & N=500 $\leftrightarrow$ N=1k & 0.00409\\\hline
\end{tabular}
\end{table}

\begin{table}[h!]
\centering
\caption{Regression metrics. The best value found for each metric is highlighted in \textbf{bold} while the second best is \underline{underlined}.}\label{tab:results}
\begin{tabular}{cllcc|ccl}
    Dataset & Model & N   & k & T & MAE ($\downarrow$) & corr ($\uparrow$) & nll ($\downarrow$) \\\hline
    \multirow{24}{*}{Solubility} 
& gpt-3.5-turbo-0125 & 700 & 1 & 0.05 & 1.120 $\pm$ 0.039 & 0.664 $\pm$ 0.043 & 7898.124 $\pm$ 3097.894 \\
& gpt-3.5-turbo-0125 & 700 & 2 & 0.05 & 0.961 $\pm$ 0.099 & 0.751 $\pm$ 0.044 & 5543.792 $\pm$ 3076.489 \\
& gpt-3.5-turbo-0125 & 700 & 5 & 0.05 & 0.880 $\pm$ 0.061 & 0.719 $\pm$ 0.031 & 11614.484 $\pm$ 10155.578 \\
& gpt-3.5-turbo-0125 & 700 & 10 & 0.05 & 0.779 $\pm$ 0.040 & 0.797 $\pm$ 0.041 & 2575.350 $\pm$ 2685.093 \\
& gpt-3.5-turbo-0125 & 700 & 5 & 0.01 & 0.872 $\pm$ 0.059 & 0.729 $\pm$ 0.059 & 3886.785 $\pm$ 4475.754 \\
& gpt-3.5-turbo-0125 & 700 & 5 & 0.1 & 0.861 $\pm$ 0.055 & 0.748 $\pm$ 0.040 & 1302.400 $\pm$ 318.958 \\
& gpt-3.5-turbo-0125 & 700 & 5 & 0.5 & 0.802 $\pm$ 0.050 & 0.796 $\pm$ 0.035 & 125.730 $\pm$ 137.078 \\
& gpt-3.5-turbo-0125 & 700 & 5 & 1.0 & 0.898 $\pm$ 0.045 & 0.758 $\pm$ 0.074 & 165.623 $\pm$ 163.004 \\
& gpt-3.5-turbo-0125 & 700 & 5 & 1.5 & 3.055 $\pm$ 1.783 & 0.230 $\pm$ 0.288 & 43.270 $\pm$ 0.430 \\
& gpt-3.5-turbo-0125 & 1 & 5 & 0.7 & 2.794 $\pm$ 0.469 & 0.050 $\pm$ 0.214 & 215.484 $\pm$ 187.379 \\
& gpt-3.5-turbo-0125 & 5 & 5 & 0.7 & 3.189 $\pm$ 0.520 & -0.042 $\pm$ 0.179 & 200.413 $\pm$ 97.635 \\
& gpt-3.5-turbo-0125 & 10 & 5 & 0.7 & 2.374 $\pm$ 0.538 & 0.298 $\pm$ 0.181 & 76.659 $\pm$ 44.318 \\
& gpt-3.5-turbo-0125 & 25 & 5 & 0.7 & 2.701 $\pm$ 0.539 & 0.156 $\pm$ 0.237 & 127.356 $\pm$ 58.155 \\
& gpt-3.5-turbo-0125 & 50 & 5 & 0.7 & 1.649 $\pm$ 0.072 & 0.412 $\pm$ 0.068 & 59.875 $\pm$ 59.008 \\
& gpt-3.5-turbo-0125 & 100 & 5 & 0.7 & 1.515 $\pm$ 0.264 & 0.484 $\pm$ 0.129 & 246.846 $\pm$ 384.317 \\
& gpt-3.5-turbo-0125 & 250 & 5 & 0.7 & 1.147 $\pm$ 0.090 & 0.656 $\pm$ 0.097 & 60.265 $\pm$ 60.187 \\
& gpt-3.5-turbo-0125 & 500 & 5 & 0.7 & 1.006 $\pm$ 0.062 & 0.696 $\pm$ 0.082 & 44.153 $\pm$ 38.803 \\
& gpt-3.5-turbo-0125 & 700 & 5 & 0.7 & 0.914 $\pm$ 0.034 & 0.727 $\pm$ 0.051 & 18.369 $\pm$ 8.975 \\
& knn & 700 & 5 & 0.7 & 1.509 $\pm$ 0.216 & 0.581 $\pm$ 0.110 & - \\
& krr & 700 & 5 & 0.7 & 0.923 $\pm$ 0.108 & 0.793 $\pm$ 0.045 & - \\
& gpr & 700 & 5 & 0.7 & 1.293 $\pm$ 0.183 & 0.571 $\pm$ 0.120 & \textbf{3.233 $\pm$ 0.006} \\
& gpt-4-0125-preview & 700 & 5 & 0.7 & \underline{0.613 $\pm$ 0.023} & \underline{0.907 $\pm$ 0.014} & 422.664 $\pm$ 305.325 \\
& gpt-4o      & 700 & 5 & 0.7 & \textbf{0.471 $\pm$ 0.030} & \textbf{0.954 $\pm$ 0.004} & \underline{17.459 $\pm$ 7.878} \\
& gpt-4o-mini & 700 & 5 & 0.7 & 0.678 $\pm$ 0.028 & 0.882 $\pm$ 0.014 & 187.039 $\pm$ 243.752 \\
    \hline 
    \multirow{24}{*}{OCM} 
& gpt-3.5-turbo-0125 & 1000 & 1 & 0.05 & 2.836 $\pm$ 0.203 & 0.447 $\pm$ 0.035 & 13046.086 $\pm$ 16116.083 \\
& gpt-3.5-turbo-0125 & 1000 & 2 & 0.05 & 2.362 $\pm$ 0.211 & 0.558 $\pm$ 0.085 & 6046.114 $\pm$ 3163.402 \\
& gpt-3.5-turbo-0125 & 1000 & 5 & 0.05 & 2.544 $\pm$ 0.145 & 0.455 $\pm$ 0.047 & 15564.800 $\pm$ 7553.086 \\
& gpt-3.5-turbo-0125 & 1000 & 10 & 0.05 & 2.545 $\pm$ 0.102 & 0.470 $\pm$ 0.039 & 14048.686 $\pm$ 5874.445 \\
& gpt-3.5-turbo-0125 & 1000 & 5 & 0.01 & 2.274 $\pm$ 0.088 & 0.507 $\pm$ 0.022 & 1328.537 $\pm$ 1502.254 \\
& gpt-3.5-turbo-0125 & 1000 & 5 & 0.1 & 2.253 $\pm$ 0.102 & 0.500 $\pm$ 0.051 & 13358.177 $\pm$ 4370.846 \\
& gpt-3.5-turbo-0125 & 1000 & 5 & 0.5 & 2.395 $\pm$ 0.134 & 0.511 $\pm$ 0.043 & 1210.133 $\pm$ 824.772 \\
& gpt-3.5-turbo-0125 & 1000 & 5 & 1.0 & 2.257 $\pm$ 0.052 & 0.551 $\pm$ 0.063 & 167.617 $\pm$ 214.227 \\
& gpt-3.5-turbo-0125 & 1000 & 5 & 1.5 & 2.795 $\pm$ 0.328 & 0.285 $\pm$ 0.151 & 62.516 $\pm$ 71.184 \\
& gpt-3.5-turbo-0125 & 1 & 5 & 0.7 & 3.868 $\pm$ 1.298 & 0.160 $\pm$ 0.077 & 51897.096 $\pm$ 103774.059 \\
& gpt-3.5-turbo-0125 & 5 & 5 & 0.7 & 2.909 $\pm$ 0.279 & 0.255 $\pm$ 0.085 & 1094.288 $\pm$ 797.185 \\
& gpt-3.5-turbo-0125 & 10 & 5 & 0.7 & 3.998 $\pm$ 0.734 & 0.264 $\pm$ 0.067 & 1542.355 $\pm$ 1845.430 \\
& gpt-3.5-turbo-0125 & 25 & 5 & 0.7 & 2.989 $\pm$ 0.156 & 0.207 $\pm$ 0.088 & 381.992 $\pm$ 647.211 \\
& gpt-3.5-turbo-0125 & 50 & 5 & 0.7 & 2.983 $\pm$ 0.127 & 0.286 $\pm$ 0.031 & 214.435 $\pm$ 159.729 \\
& gpt-3.5-turbo-0125 & 100 & 5 & 0.7 & 3.093 $\pm$ 0.064 & 0.287 $\pm$ 0.031 & 305.642 $\pm$ 229.304 \\
& gpt-3.5-turbo-0125 & 250 & 5 & 0.7 & 2.931 $\pm$ 0.296 & 0.374 $\pm$ 0.074 & 117.576 $\pm$ 45.203 \\
& gpt-3.5-turbo-0125 & 500 & 5 & 0.7 & 2.656 $\pm$ 0.172 & 0.411 $\pm$ 0.062 & 143.912 $\pm$ 61.103 \\
& gpt-3.5-turbo-0125 & 1000 & 5 & 0.7 & 2.219 $\pm$ 0.137 & 0.555 $\pm$ 0.048 & 139.080 $\pm$ 30.690 \\
& knn & 1000 & 5 & 0.7 & 2.171 $\pm$ 0.194 & 0.586 $\pm$ 0.036 & - \\
& krr & 1000 & 5 & 0.7 & \underline{1.934 $\pm$ 0.081} & \textbf{0.723 $\pm$ 0.029} & - \\
& gpr & 1000 & 5 & 0.7 & 2.312 $\pm$ 0.122 & 0.506 $\pm$ 0.058 & \textbf{3.265 $\pm$ 0.005} \\
& gpt-4-0125-preview & 1000 & 5 & 0.7 & 2.053 $\pm$ 0.119 & 0.631 $\pm$ 0.021 & 1931.768 $\pm$ 758.246 \\
& gpt-4o      & 1000 & 5 & 0.7 & \textbf{1.863 $\pm$ 0.151} & 0.649 $\pm$ 0.060 & 26.588 $\pm$ 19.877 \\
& gpt-4o-mini & 1000 & 5 & 0.7 & 2.102 $\pm$ 0.096 & 0.552 $\pm$ 0.039 & 815.057 $\pm$ 331.204 \\
& gemini-2.5-flash & 1000 & 5 & 0.7 & 2.050 $\pm$ 0.125 & \underline{0.624 $\pm$ 0.040} & \underline{12.492 $\pm$ 1.693} \\
\end{tabular}
\end{table}

\clearpage
\subsection{Solubility}\label{sec:solubility}

\paragraph{Solubility} In this study, we considered three datasets. First, to evaluate BO-ICL, we applied it to the ESOL\cite{Delaney2004-de} dataset, using IUPAC names as representations for molecules and measured LogS value labels. This solubility dataset is a benchmark largely used to evaluate models and was employed to allow broad comparison with the literature. 

\begin{figure}[h!]
    \centering
    \includegraphics[width=0.8\linewidth]{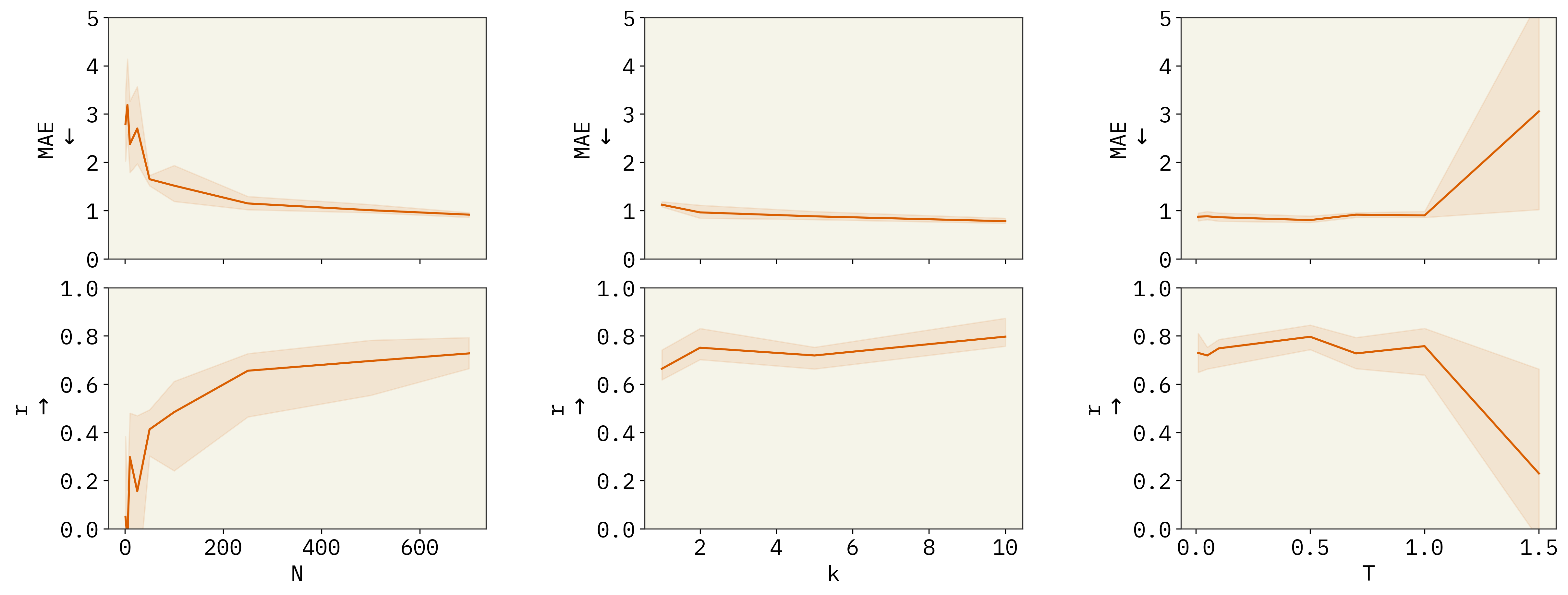}
    \caption{Performance metrics for hyperparameter tuning on the solubility dataset. The model \texttt{gpt-3.5-turbo-0.125} shows consistent improvement as the number of samples in the model's memory (N) increases. However, its performance is relatively insensitive to the number of in-context examples (k) and the temperature (T). Only at high temperature values does performance drop significantly, likely due to increased hallucinations.}
    \label{fig:sol-metrics}
\end{figure}

\begin{figure}[h!]
    \centering
    \includegraphics[width=\linewidth]{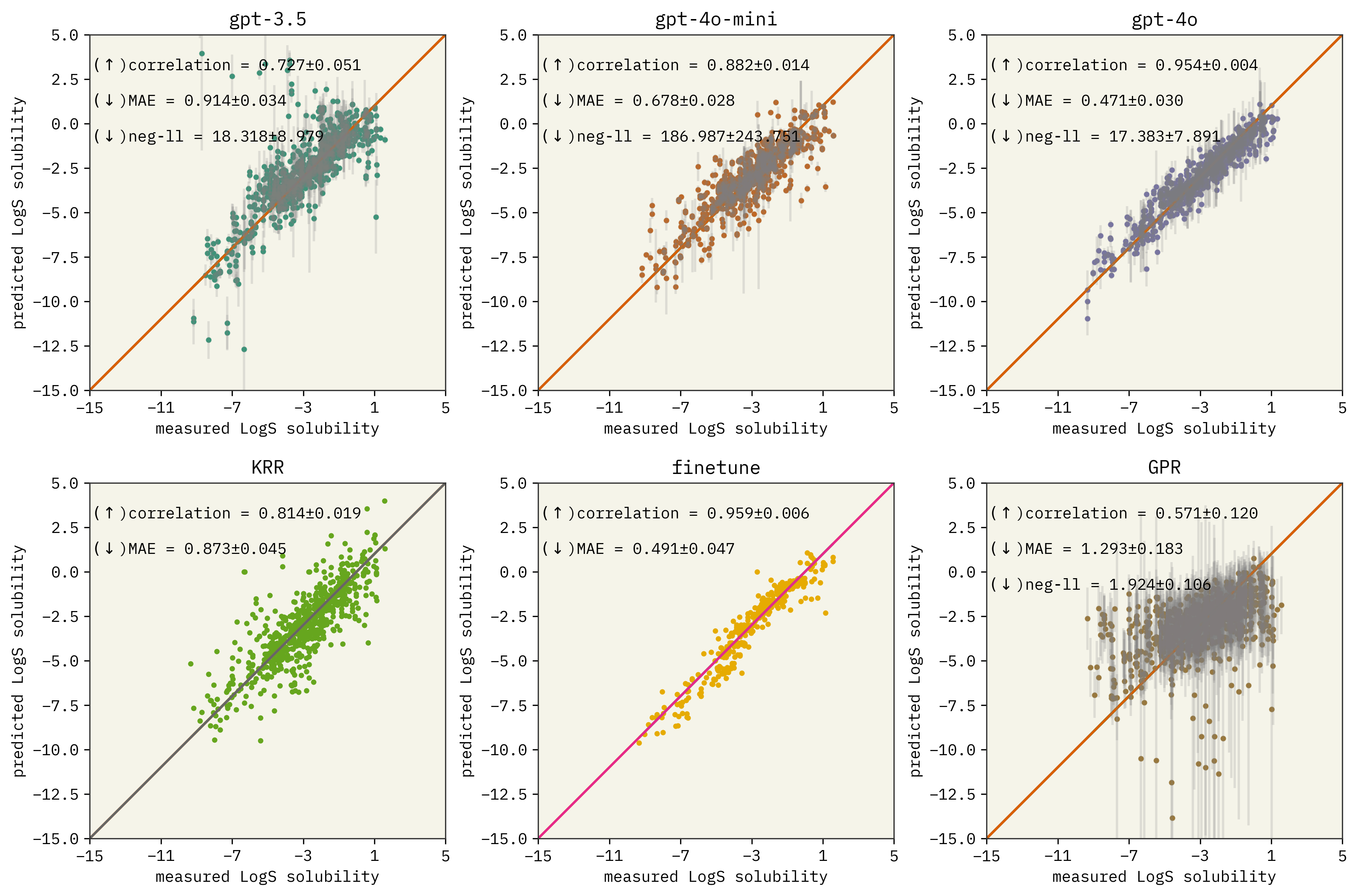}
    \caption{Parity plots for the regression on the solubility dataset task across different models. Each model was evaluated over five independent replicates, with each plot aggregating all predicted vs. true values. Reported metrics reflect the mean and standard deviation across replicates. }
    \label{fig:sol-models}
\end{figure}

\begin{figure}[h!]
    \centering
    \includegraphics[width=\linewidth]{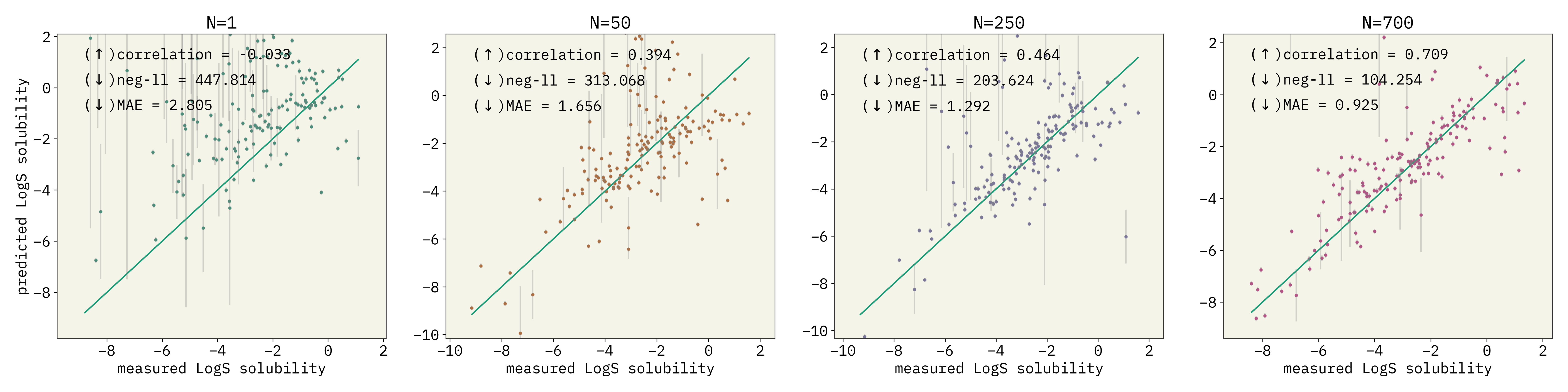}
    \includegraphics[width=\linewidth]{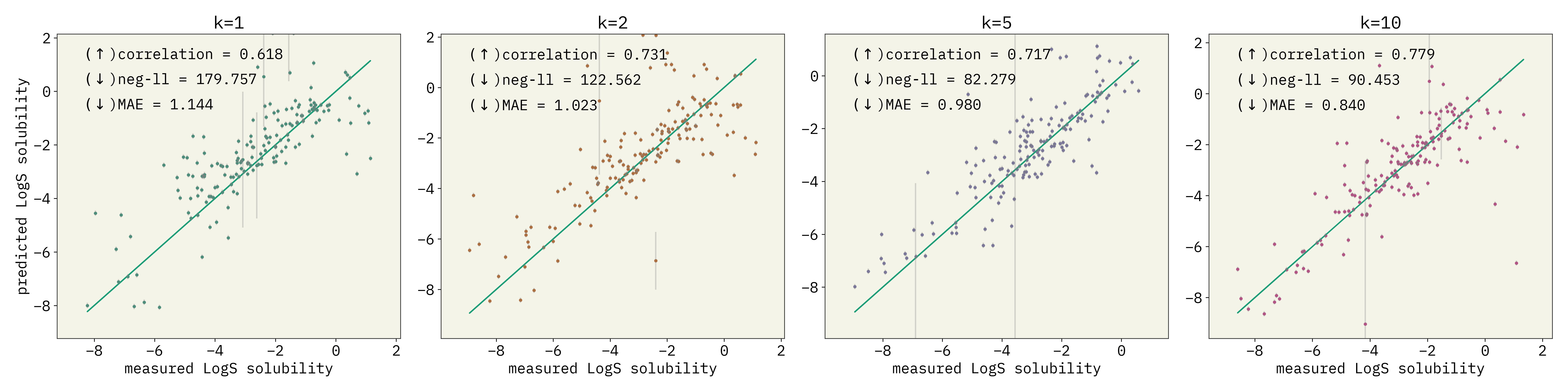}
    \includegraphics[width=0.7\linewidth]{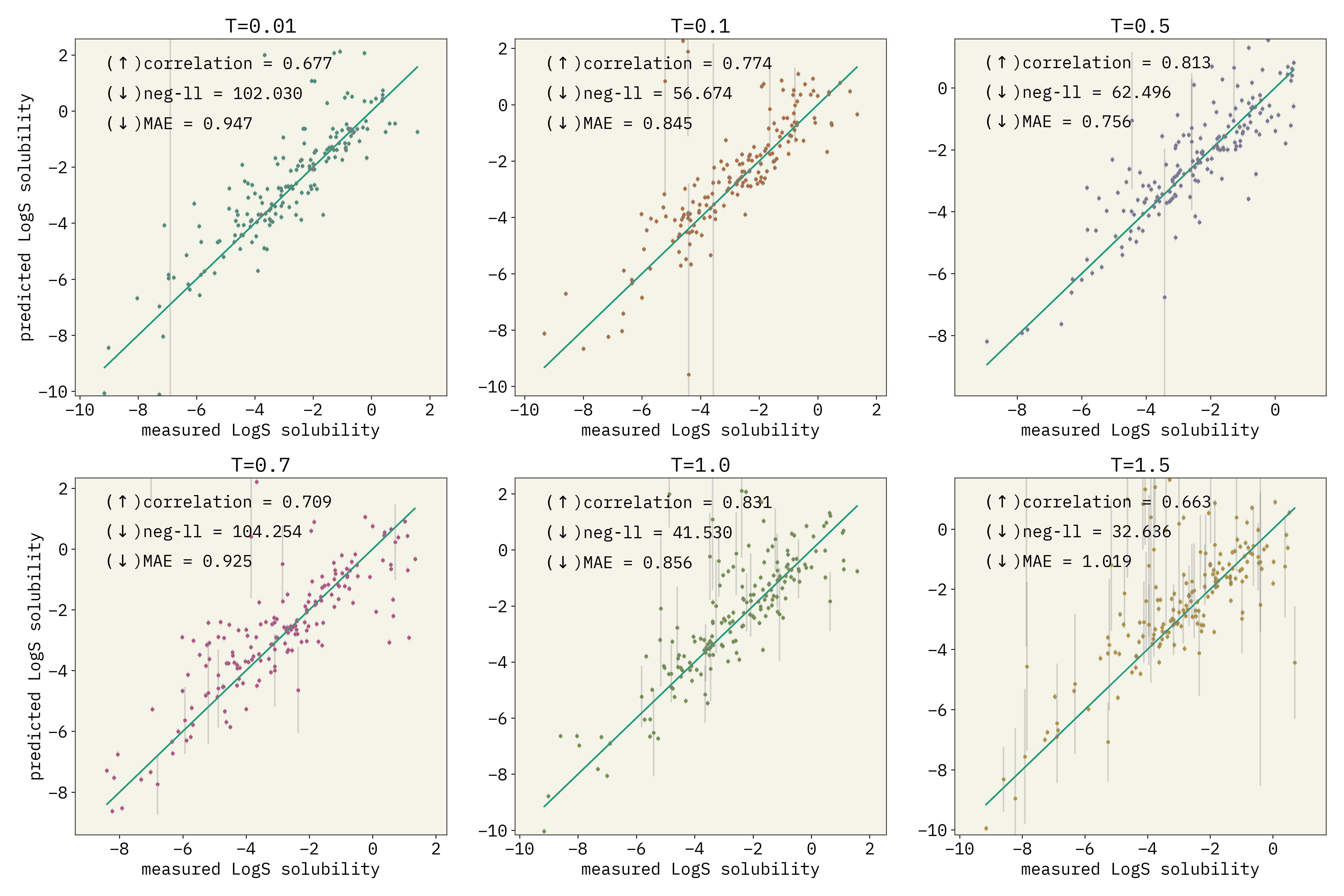}
    \caption{Illustrative parity plots from the hyperparameter tuning experiment on the solubility. Each inset title indicates the hyperparameter being varied. Unless specified otherwise in the title, the default configuration used is N = 700, k = 5, and T = 0.7.}
    \label{fig:sol-hypers}
\end{figure}

\clearpage
\subsection{Regression - OCM}\label{sec:regocm}

\begin{figure}[h!]
    \centering
    \includegraphics[width=0.8\linewidth]{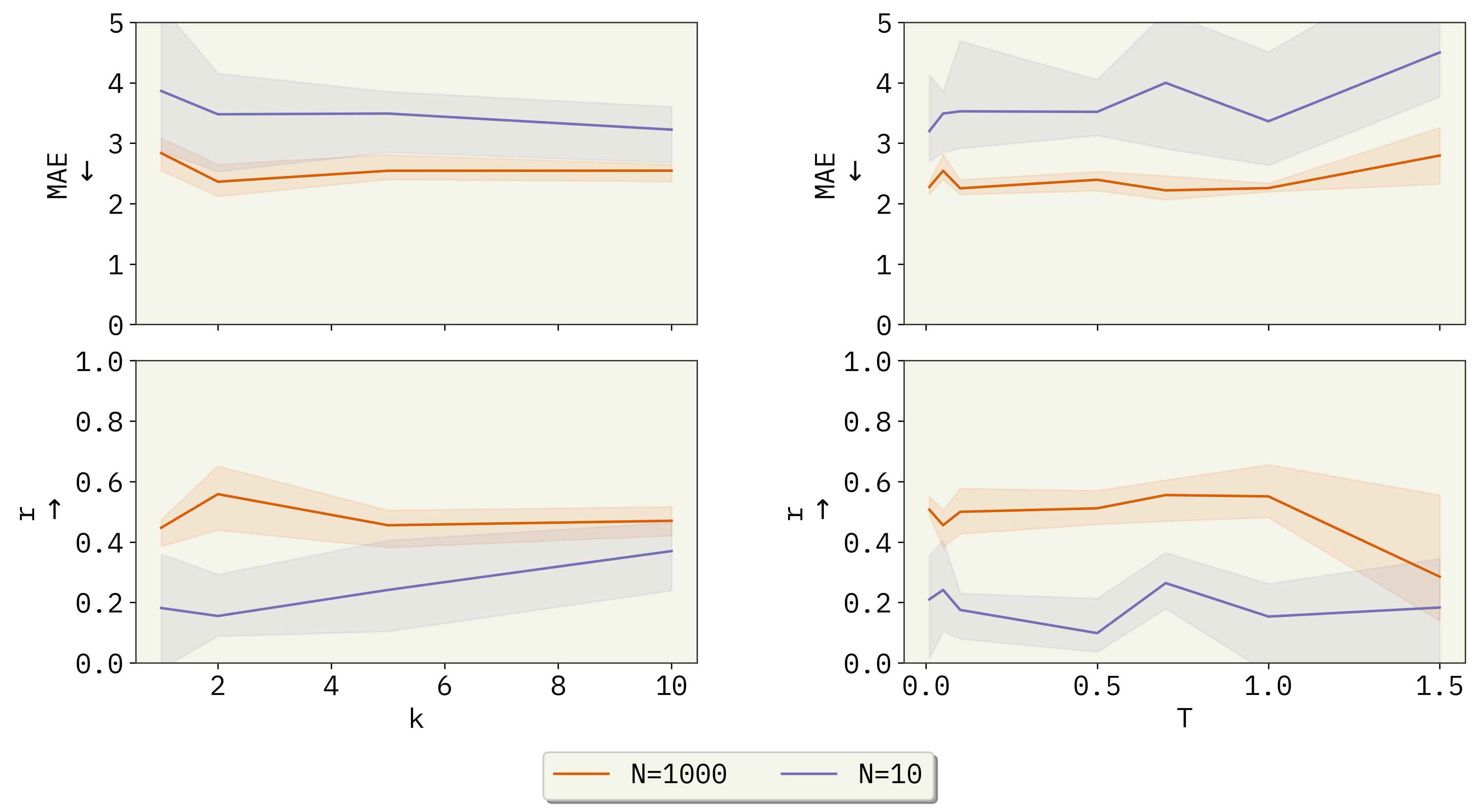}
    \caption{To assess the impact of hyperparameters in low-data scenarios, we evaluated performance under two conditions: N = 1000 and N = 10. In the low-data regime (N = 10), the number of in-context examples (k) plays a more critical role, with performance steadily improving as k increases. In contrast, the effect of temperature (T) remains consistent with previous findings on the solubility dataset: model performance is largely stable across T values, except at high temperatures (T = 1.5), where a notable drop occurs due to increased hallucinations.}
    \label{fig:ocm-metrics2}
\end{figure}

\begin{figure}[hbt!]
    \centering
    \includegraphics[width=\linewidth]{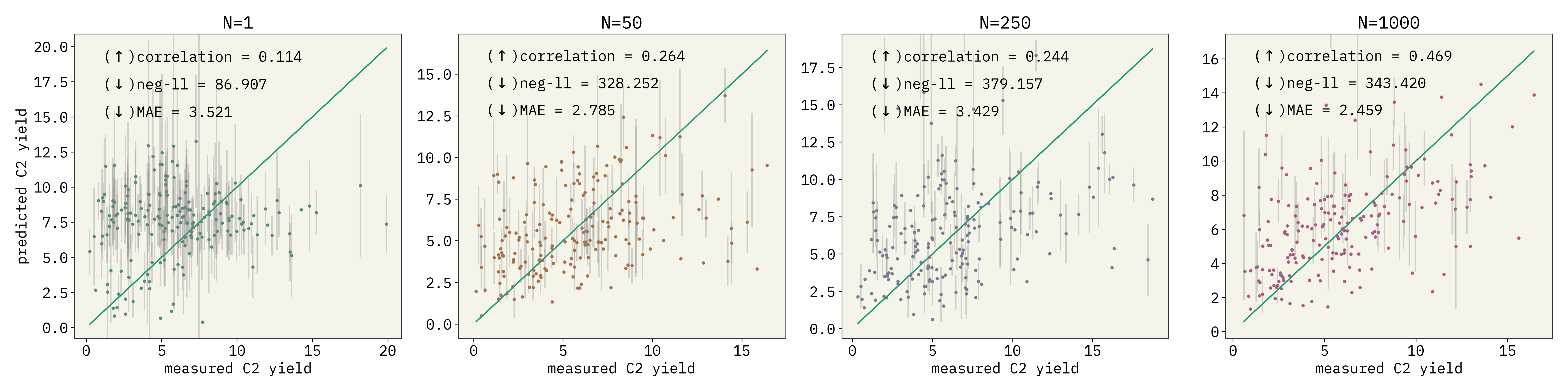}
    \includegraphics[width=\linewidth]{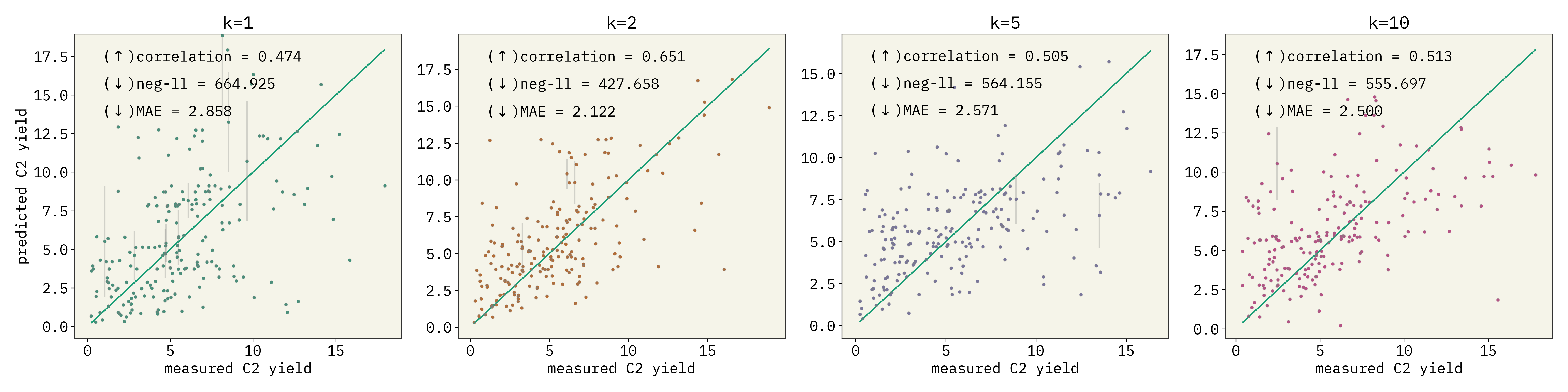}
    \includegraphics[width=0.7\linewidth]{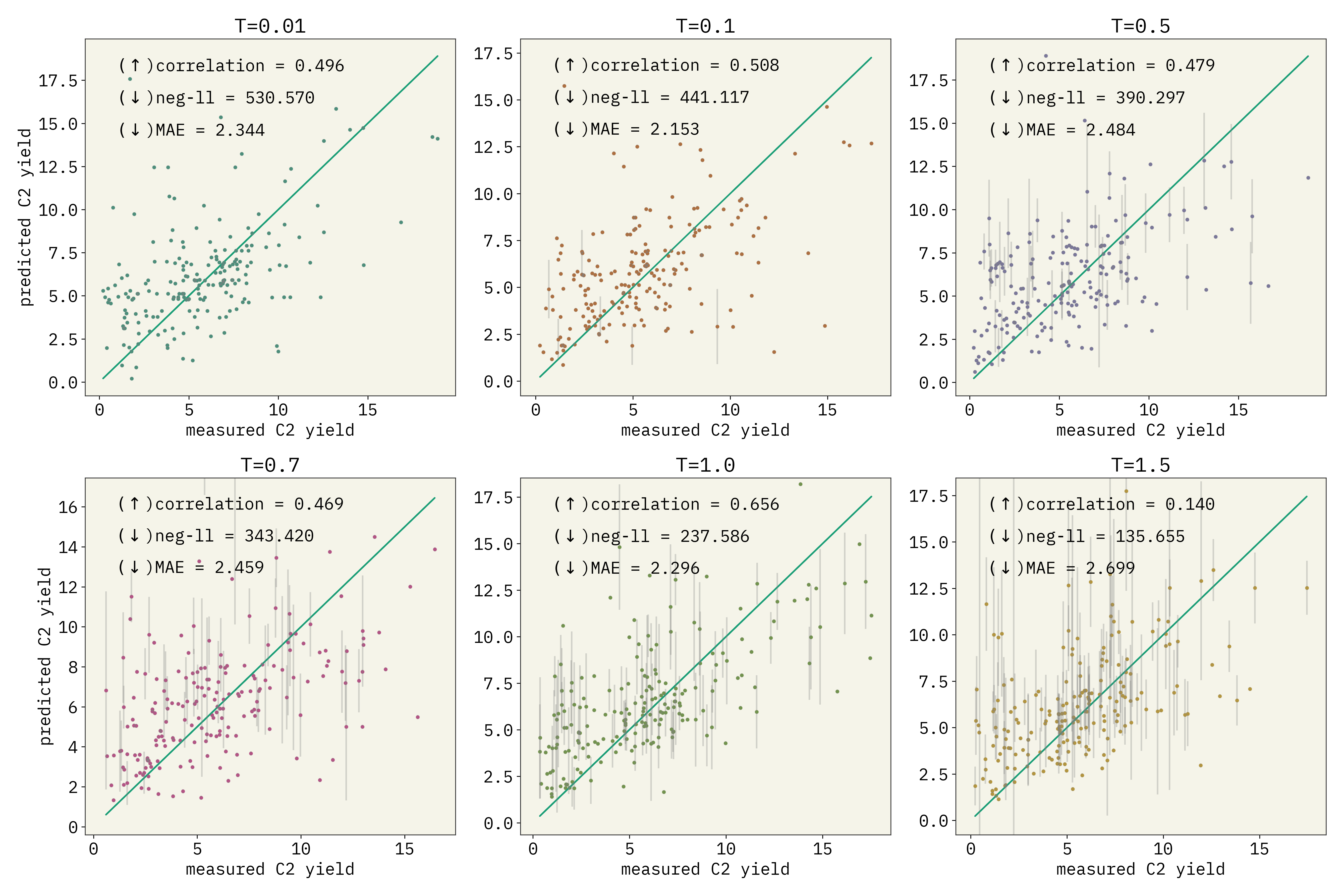}
    \caption{Illustrative parity plots from the hyperparameter tuning experiment on the OCM dataset. Each inset title indicates the hyperparameter being varied. Unless specified otherwise in the title, the default configuration used is N = 1000, k = 5, and T = 0.7.}
    \label{fig:ocm-hypers}
\end{figure}

\clearpage
\subsection{MMR Dependence}\label{sec:mmr}

\begin{figure}[hbt!]
    \centering
    \includegraphics[width=\linewidth]{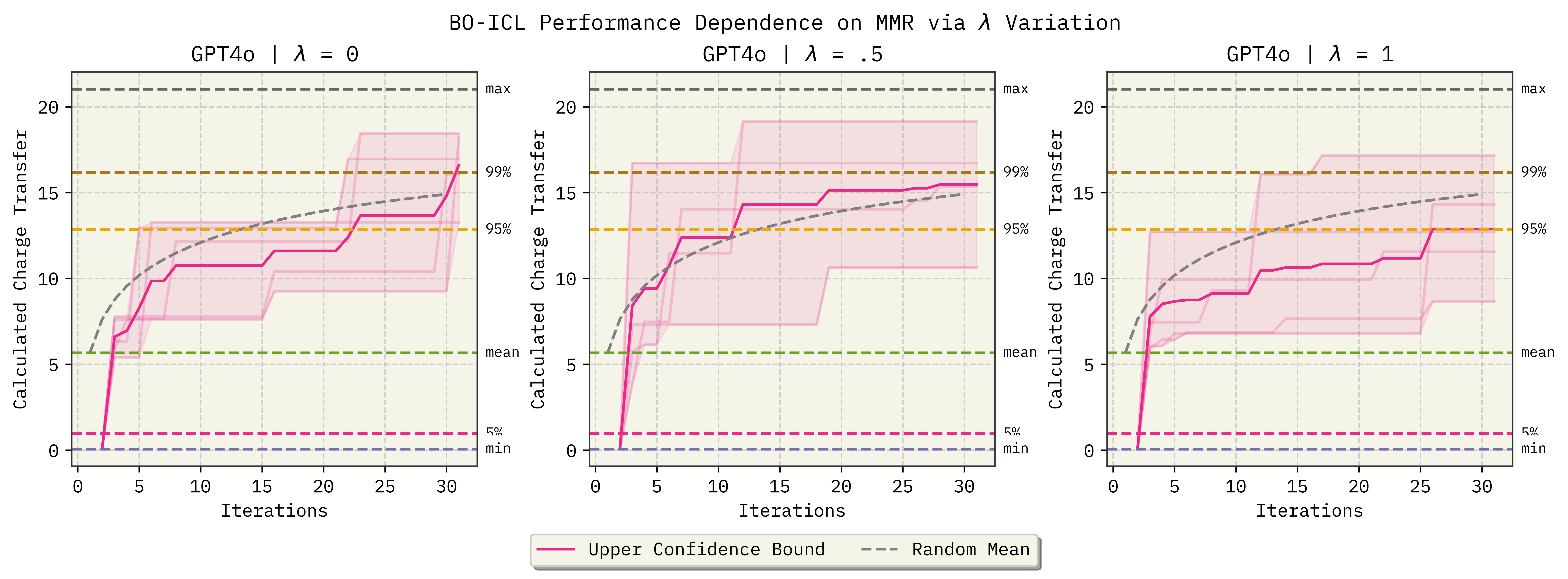}
    \caption{OCM dataset with $\lambda$ variation. At $\lambda$ is equal to zero, query relevance is ignored and $\lambda$ equal to 1 mimics sampling using cosine-similarity in that the the closest samples to the reference populate the sub-pool.}
    \label{fig:MMR_lambda}
\end{figure}

We see that a balance between diversity and query-relevant examples significantly impacts performance and thus justifies the use of MMR at a user selected lambda.(Figure ~\ref{fig:MMR_lambda})

